\documentclass[useAMS,usenatbib,usegraphicx]{mn2e}
\usepackage{psfig}   
\usepackage{graphicx}
\usepackage{subfigure}

\title[Stochasticity in binary destruction]{The same, but different: 
Stochasticity in binary destruction}

\author[R.~J.~Parker \& S.~P.~Goodwin]{
  Richard J.~Parker$^{1}$\thanks{E-mail: rparker@phys.ethz.ch} and Simon P.~Goodwin$^{2}$
  \vspace*{0.1cm}\\
   $^1$ Institute for Astronomy, ETH Z{\"u}rich, Wolfgang-Pauli-Strasse 27, 8093, Z{\"u}rich, 
   Switzerland \\
   $^2$ Department of Physics and Astronomy, University of Sheffield, Sheffield, S3 7RH, UK}

\begin{document}

\date{}
                             
\pagerange{\pageref{firstpage}--\pageref{lastpage}} \pubyear{2012}

\maketitle

\label{firstpage}

\begin{abstract}
Observations of binaries in clusters tend to be of visual 
binaries with separations of 10s -- 100s~au.  Such binaries are
`intermediates' and their destruction or survival depends on the
exact details of their individual dynamical history.  We 
investigate the stochasticity of the
destruction of such binaries and the differences between the initial
and processed populations using $N$-body simulations.  We 
concentrate on Orion Nebula 
Cluster-like clusters, where the observed binary separation 
distribution ranges from 62 -- 620\,au.

We find that, starting from the same initial binary
population in statistically identical clusters, the number 
of intermediate binaries that are destroyed after 1\,Myr 
can vary by a factor of $>2$, and that the resulting separation
distributions can be statistically completely different in initially substructured clusters.  We also find
that the mass ratio distributions are altered (destroying more low
mass ratio systems), but not as significantly as the binary fractions
or separation distributions.  We conclude that finding very different
intermediate (visual) binary populations in different clusters does
not provide conclusive evidence that the initial populations were different.
\end{abstract}


\section{Introduction}

The nature of star formation is one of the great unsolved problems in
astrophysics. The formation of stars is extremely interesting in
itself, but also has implications for galaxy formation and evolution,
and planet formation. In recent years, studies of young
star forming regions have shown that the initial mass function (IMF)
is invariant, at least on nearby  galactic scales \citep*{Bastian10}.

It is unclear whether this apparent universality of star formation in
the IMF is also mirrored in the primordial binary population. Most
stars form in binaries \citep{Goodwin05a,Kroupa08}, but the  picture
is clouded by subsequent dynamical evolution in some clustered
environments \citep[e.g.][]{Kroupa95a,Kroupa95b,Parker09a,Parker11c},
making it difficult to conclude whether or not binary  formation in
different star forming regions is also universal \citep{King12a}. 

By comparing the results of $N$-body simulations to  observations of
binaries in both clusters and the Galactic field it is possible to
account for this dynamical evolution and then infer the probable
initial conditions of star formation, a process known as ``reverse
engineering'' or ``inverse population synthesis''
\citep{Kroupa95a}. For this purpose the results of many simulations
($\geq 10$) are usually averaged together to obtain a 1--$\sigma$
uncertainty, and then compared to observations.

Most observations of the binary separation distribution in young
clusters tend to probe the visual separation regime
\citep[e.g.][]{Patience02,Reipurth07,King12a}, in which binaries
typically  have separations between several tens, to several hundreds
of au (this depends on distance and cluster surface density).

Taking the Orion Nebula Cluster (ONC) as an example,  the
observations probe the separation range 62 -- 620\,au
\citep{Reipurth07}.  Binaries with shorter separations are 
difficult to detect in
clusters, whereas those with wider separations become
indistinguishable against the background  of other cluster members 
\citep[if they even exist in such clusters, ][]{Scally99,Parker09a}.  Comparison with averaged numerical simulations
\citep{Parker09a,Parker11c} suggest good agreement with a primordial
field-like separation distribution and an initial binary fraction of around
75\,per cent.  

Unfortunately, as we will investigate in detail in this
paper, this 10s to 100s~au `intermediate' binary separation range is one which is 
affected stochastically by dynamical interactions.

\citet{Heggie75} and \citet{Hills75a} investigated the dynamically processing
of binaries.  They divided binaries into two broad classes:  hard, and soft.  Hard binaries have a binding energy that exceeds the local Maxwellian energy of stars in the cluster and are so tightly bound that it is
extremely rare for an encounter to destroy them (indeed, encounters
tend to extract energy, making them harder).  Soft binaries have a binding energy that is less than local Maxwellian energy and  are so
loosely bound that single distant encounters, or even the tidal field
of a cluster, can destroy them \citep[although they are so easy to make that
a transient population can exist, see][]{Moeckel11}.

Based on these definitions, it is also possible to define a third dynamical class of binaries: intermediate binaries. Intermediate binaries are those inbetween hard and soft  (their binding energy is comparable to the local Maxwellian energy of stars in the cluster), whereby a
single relatively close encounter, or several distant 
encounters {\em may} destroy them.  Therefore if an intermediate binary survives
depends on the exact details of its dynamical history and an element
of `luck' in the number and severity of encounters it has.

In a smooth, spherical system such as a Plummer sphere, the boundary between hard and 
soft binaries, $a_{\rm hs}$, can be estimated following \citet{Binney87}, as
\begin{equation}
a_{\rm hs} = \frac{3}{2}\frac{r_{1/2}}{N_{\rm sys}},
\label{hard_soft}
\end{equation}
where $r_{1/2}$ is the half-mass radius of the cluster, and $N_{\rm sys}$ is the number of stellar systems in the cluster. Adopting the 
current half-mass radius of the ONC as 0.8\,pc \citep{Hillenbrand98}, and the number of stars as $\sim$~1500 \citep{King12a}, then 
$a_{\rm hs} \simeq 250$\,au. 

However, the hard-soft boundary is not a sharp boundary.  Destruction
depends not only on the typical encounter energy/velocity, but also on
having an encounter, and hence an element of `luck' in having or
avoiding a destructive encounter.  The encounter timescale depends on
density, but with (as we will show) a stochastic 
element\footnote{In the Galactic field the hard-soft boundary is formally much lower than in clusters as the velocity dispersion is much higher than in clusters.  However, many formally soft binaries can survive for Gyr as the encounter timescale is so long.}.  
Binaries a factor of 2 or 3 above $a_{\rm hs}$ can survive if they avoid strong encounters, and binaries a factor of 2 or 3 below $a_{{\rm hs}}$ can be destroyed.

As the hard-soft boundary depends on {\em local} density it varies radially in smooth distributions such as Plummer spheres. It is also a very difficult quantity to define in substructured distributions such as fractals as the density can vary significantly. 

In this paper we investigate the consequences of the stochasticity of
intermediate binary destruction in star clusters.  This is
particularly important because, as we have discussed, observations generally
cover the intermediate binary population.  We  
evolve a variety of clumpy and smooth clusters 
containing exactly the same initial binary population
(identical primary and secondary masses, semi-major axes, and
eccentricities) and examine the intermediate binary population
after 1~Myr.  We describe the simulation set-up in
Section~\ref{method}, we present our results  in
Section~\ref{results}, we provide a discussion in
Section~\ref{discuss} and we conclude in Section~\ref{conclude}. 

\section{Method}
\label{method}

In this Section we describe the method used to set up and run the
numerical simulations of our model clusters.

\subsection{Binary population}
\label{binary_setup}

We set the clusters up with only one primordial binary
population. This enables an investigation into the effects of
morphology and dynamics  on a constant initial separation distribution
of intermediate binaries to compare to the observational data. 

Earlier work has shown that in a dense ONC-like cluster, a primordial
binary population will be affected by dynamical interactions, which
both lowers  the primordial binary fraction and alters the initial
semi-major axis (hereafter separation) distribution 
\citep[e.g.][]{Kroupa99,Parker09a,Parker11c}. 

Recently, \citet{King12a} have placed observational and theoretical
constraints on the primordial binary fraction and separation
distribution in the ONC, and find that a G-dwarf field-like  separation
distribution \citep{Duquennoy91,Raghavan10}, and an initial binary
fraction of $\sim$75\,per cent (also confirmed from theoretical
considerations by \citet{Kaczmarek11} and \citet{Parker11c})
represents the most likely primordial binary population.

In this work we adopt an initial binary fraction of 100\,per cent, and
a field-like separation distribution
\citep{Duquennoy91,Fischer92,Raghavan10}. As our clusters  are
relatively dense initially, the widest binaries in the field-like
separation distribution are not physically bound \citep{Parker09a},
and the starting binary fraction in the simulations is closer to 75\,per cent. 

We draw the primary masses from a \citet{Kroupa02} IMF of the form
\begin{equation}
 N(M)   \propto  \left\{ \begin{array}{ll} M^{-1.3} \hspace{0.4cm} m_0
   < M/{\rm M_\odot} \leq m_1   \,, \\ M^{-2.3} \hspace{0.4cm} m_1 <
   M/{\rm M_\odot} \leq m_2   \,,
\end{array} \right.
\end{equation}
where $m_0$ = 0.1\,M$_\odot$, $m_1$ = 0.5\,M$_\odot$, and  $m_2$ =
50\,M$_\odot$. We do not include brown dwarfs in the
simulations as these are not present in the observational samples with
which we will compare our simulations. Secondary masses are drawn  from a flat mass ratio
distribution, in accordance with observations of the distribution in
the Galactic field \citep{Reggiani11}.  However, we limit the lower
mass of a companion to be 0.1\,M$_\odot$; this means that lower-mass
stars do not have a full range of mass ratios. For example, a 0.15\,M$_\odot$
primary can only have companions in the range 0.1 -- 0.15\,M$_\odot$.
If a companion of mass $<$ 0.1\,M$_\odot$ is selected we draw a new
random mass ratio until a companion $\geq$ 0.1\,M$_\odot$ is selected.

In accordance with observations of the field, we select binary periods
from the log-normal fit to the G-dwarfs in the field by
\citet{Duquennoy91} -- see also \citet{Raghavan10},  which has also
been extrapolated to fit the period distributions of the K- and
M-dwarfs \citep{Mayor92,Fischer92}:
\begin{equation}
f\left({\rm log_{10}}P\right)  \propto {\rm exp}\left \{ \frac{-{({\rm
      log_{10}}P - \overline{{\rm log_{10}}P})}^2}{2\sigma^2_{{\rm
      log_{10}}P}}\right \},
\end{equation}
where $\overline{{\rm log_{10}}P} = 4.8$, $\sigma_{{\rm log_{10}}P} =
2.3$ and $P$ is  in days. We convert the periods to semi-major axes
using the masses of the binary components.

The eccentricities of binary stars are drawn from a thermal
distribution \citep{Heggie75,Kroupa08} of the form
\begin{equation}
f_e(e) = 2e.
\end{equation}
In the sample of \citet{Duquennoy91}, close binaries (with periods
less than 10 days) are almost exclusively on tidally circularised
orbits. We account for this by reselecting the eccentricity of a
system if it exceeds the following  period-dependent
value\footnote{\citet{Kroupa95b} and \citet{Kroupa08} provides a more
  elaborate `eigenevolution' mechanism to incorporate interactions
  between the primary star and its protostellar disk during tidal
  circularisation.  However, this mechanism also alters the mass ratio
  distribution, causing a deviation from the flat mass ratio
  distribution observed in the Galactic field \citep{Reggiani11}.}:
\begin{equation}
e_{\rm tid} = \frac{1}{2}\left[0.95 + {\rm tanh}\left(0.6\,{\rm
    log_{10}}P - 1.7\right)\right].
\end{equation}

We combine the primary and secondary masses of the binaries with their
semi-major axes and eccentricities to determine the relative velocity
and radial components of the stars in  each system. The binaries are
then placed at the centre of mass and velocity for each system in
either the fractal distribution or Plummer sphere (see
Section~\ref{morph}). 

Note that the exact details of the initial binary distribution do not
matter.  The following results would be true of {\em any} initial 
distribution of intermediate binaries in a cluster.

\subsection{Cluster morphologies}
\label{morph}

We set up clusters containing 1500 stars (i.e.\,\,750 binary systems),
and adopt two different morphologies. Firstly, we create fractal
clusters \citep{Cartwright04,Goodwin04a} to create clusters with
substructure, and secondly, we use Plummer spheres \citep{Plummer11}
to enable a comparison between centrally concentrated, smooth
clusters,  and the substructured clusters. 

\subsubsection{Fractal clusters}

Observations of young, dynamically unevolved star forming regions
indicate that a large amount of substructure is present
\citep[e.g.][]{Cartwright04,Sanchez09}. The  most convenient way of
describing substructure is via the fractal, in which the amount of
substructure is set by just one number, the fractal dimension, $D$
\citep{Goodwin04a}.  We adopt a moderate amount of substructure ($D =
2.0$). 

The velocities of systems in the fractal are drawn from a Gaussian of
mean zero, and the fractal is constructed in such a way that nearby
stars have similar velocities, whereas the  velocities of distant
stars can be very different \citep[see][for a more detailed
  description]{Goodwin04a,Parker11c}. The initial radius of the
fractal is 1\,pc, and we scale the velocities  so the cluster has a
virial ratio $Q = 0.3$, which is subvirial or `cool'. These initial
conditions have been successful in explaining the level of mass
segregation in the ONC  through dynamics \citep{Allison09b,Allison10},
and can account for the formation of Trapezium-like systems
\citep{Allison11}. 

\subsubsection{Plummer spheres}

No two fractals are identical and to the eye two statistically
identical fractals can look very different.  It is therefore
desirable to test whether any differences in the intermediate
separation distribution are not simply due to the exact details of the
fractal realisations. We therefore conduct simulations in which 
we evolve the same primordial binary population in a radially 
smooth, centrally concentrated Plummer sphere \citep{Plummer11}.
Whilst no two Plummer spheres are identical, their initial structures
and their evolution are much more similar than fractals.

The positions and velocities of the systems are determined according
to the prescription in \citet*{Aarseth74}. We construct Plummer
spheres with an initial half-mass radius  $r_{1/2} = 0.1$\,pc 
  (corresponding to a hard-soft boundary of $\sim$70\,au), and set
them to be in virial equilibrium initially ($Q = 0.5$).\\

The binaries are then randomly  assigned a system position and
velocity in the fractal or Plummer sphere, which varies with each
realisation of the cluster morphology. We run 10 realisations of each
morphology, identical apart from the random number seed used to
initialise the positions and velocities of the systems. In each
cluster  we place the same population of binary stars (see Section
\ref{binary_setup}). We do not include stellar evolution in the
simulations. The simulations are run for 1\,Myr  using the
\texttt{kira} integrator in the Starlab package
\citep[e.g.][]{Zwart99,Zwart01}.

\subsection{Summary}

To summarise: we take a single initial binary population, always the
same in every way, and place it in ten realisations of a fractal
cluster and ten realisations of a Plummer sphere.  We evolve each
cluster for 1~Myr and examine the remaining intermediate binary
population in each cluster.

We have chosen to compare our simulations with the ONC.  Firstly, 
there are good observations of visual binaries in the ONC 
in the separation range 62 -- 620\,au \citep{Reipurth07}.  Secondly,
the density of the ONC suggests that the hard-soft boundary 
lies within this separation range \citep{Kroupa99,Parker09a}\footnote{In this paper we will keep the separation range with which we compare the simulations fixed, but we note that 
our two different suites of simulations may have quite different hard-soft boundaries. For example, our Plummer sphere clusters reach higher densities (and therefore contain more soft binaries) than the fractal clusters.}.
Finally, the ONC has a large enough population ($N \sim 1500$ stars) that we
have a significant population in this separation range in each
cluster.  In later papers we will discuss other separation ranges,
different cluster masses and the
effects of small number statstics, but for now we will concentrate on
the currently observed intermediate binary population in a fairly
massive ONC-like cluster.

\section{Results}
\label{results}

\begin{table*}
\caption[bf]{The numbers of binaries in the separation range 62--620\,au in the fractal cluster simulations at 0\,Myr (first row) and at 1\,Myr (second row). Each 
simulation has 106 binaries initially.}
\begin{center}
\begin{tabular}{|c|c|c|c|c|c|c|c|c|c|c|}
\hline 
simulation & a & b & c & d & e & f & g & h & i & j  \\
\hline
$N_{\rm bin;  0\,Myr}$ & 106 & 106 & 106 & 106 & 106 & 106 & 106 & 106 & 106 &  106 \\
\hline
$N_{\rm bin;  1\,Myr}$ & 87 & 83 & 67 & 74 & 73 & 77 & 66 & 81 & 62 & 69 \\
\hline
\end{tabular}
\end{center}
\label{fractal_simulations}
\end{table*}

\begin{table*}
\caption[bf]{The numbers of binaries in the separation range 62--620\,au in the Plummer sphere cluster simulations at 0\,Myr (first row) and at 1\,Myr 
(second row). Not all 106 binaries are physically bound at the start of each simulation due to the high initial densities of the Plummer spheres.}
\begin{center}
\begin{tabular}{|c|c|c|c|c|c|c|c|c|c|c|}
\hline 
simulation & a & b & c & d & e & f & g & h & i & j  \\
\hline
$N_{\rm bin; 0\,Myr}$ & 102 & 100 & 102 & 103 & 105 & 104 & 104 & 101 & 104 & 102 \\
\hline
$N_{\rm bin; 1\,Myr}$ & 48 & 39 & 61 & 42 & 54 & 65 & 52 & 50 & 64 & 54 \\
\hline
\end{tabular}
\end{center}
\label{Plummer_simulations}
\end{table*}

Our {\em initial} binary population is formed with 106 binaries 
in the range 62 -- 620\,au, and this is the initial population for
{\em every} cluster.  Clusters differ only in the random number seeds
which set the system positions and velocities, not the system properties.

In this Section, we will first examine the differences in the numbers of intermediate binaries which are destroyed after 1\,Myr, before turning 
our attention to the separation distributions and the mass ratio distributions of these binaries.

\subsection{Binary fraction}

In Table~\ref{fractal_simulations} we present the initial (0\,Myr) and final (1\,Myr) numbers of binaries in the separation range 62--620\,au 
in each of the ten fractal cluster simulations. Each cluster has 106 binaries in this range initially, but the final number of binaries varies significantly, with 
extrema of 62 and 87 binaries (simulations i and a, respectively). Thus between 42\,per cent and 18 per cent of the initial population has been destroyed. Looking just at this range, we started with 
212 stars in 106 systems (a binary fraction of unity\footnote{We define the binary fraction, $f_{\rm bin} = \frac{B}{S + B}$, where $B$ is the number of binaries 
(and higher order multiple systems), 
and $S$ is the number of singles.}) and the extremes after 1~Myr are 212 stars in 150 systems (a binary fraction of 0.41) and 212 stars in 125 systems (a binary 
fraction of 0.70). 

Turning to the Plummer sphere clusters, in Table~\ref{Plummer_simulations} we show the initial (0\,Myr) and final (1\,Myr) numbers of binaries in the 
separation range 62--620\,au for each of our ten simulated clusters. There is a slight variation in the initial number of binaries detected by our algorithm 
(and no cluster has its full compliment of 106 initial binaries identified), due to the high initial densities of the Plummer spheres. 

In the two most extreme cases, 39 binaries remain in a cluster that contained 100 initially (simulation b), and 65 binaries remain in a cluster that contained 
105 initially (simulation f).  Thus between 71\,per cent and 37\,per cent of the initial population has been destroyed. In terms of the binary fraction, we started 
with 212 stars in 112 systems (a binary fraction of 0.89 -- simulation b), and 212 stars in 108 systems (a binary fraction of 0.96 -- simulation f). After 1\,Myr 
the binary fractions in these clusters are 0.23 and 0.44, respectively. 

The {\em total} binary fraction in the cluster of course also depends on the numbers of systems with separations outside 
of this range that have been destroyed. 

More binaries are destroyed in the Plummer sphere clusters than the fractal clusters.  The reason for this is that we produce an intermediate binary
(say of separation 500~au) and place it at random within the
simulation.  If the binary is placed in a low-density region where the
typical separation between stars is, say, 3000~au, then it is clearly
identified as a binary system.  However, if a 500~au binary is placed in a
dense region with a typical inter-star separation of, say, 800~au,
then it is no longer a `binary'.  The handful of the 106 intermediate
separation systems placed near the centre of a Plummer sphere are
therefore `destroyed' at time zero.

The localised substructure in the fractal clusters is not as dense as the central regions of the Plummer spheres, and so all binaries that `form' in the fractals remain physically bound. 
The maxiumum densities in the fractal clusters are around 1000 --
2000\,M$_\odot$~pc$^{-3}$, 
whilst the central densities of the Plummer
spheres are around $6 - 7 \times 10^4$\,M$_\odot$~pc$^{-3}$.  Therefore, 
as the simulation progresses there will be significantly more and
closer encounters in the centres of the Plummer spheres, which process
the intermediate binaries more than in fractal clusters.  We noted
above that after 1~Myr the 106 intermediate binaries in the fractal
clusters had been reduced to between 62 and 87 systems.  In the much
denser Plummer spheres the final numbers of systems are between 39 and
65 -- a far more destructive environment.

\begin{figure*}
  \begin{center}
\setlength{\subfigcapskip}{10pt}
\vspace*{-0.2cm}
\hspace*{-0.5cm}\subfigure[]{\label{rei_sep-a}\rotatebox{270}{\includegraphics[scale=0.26]{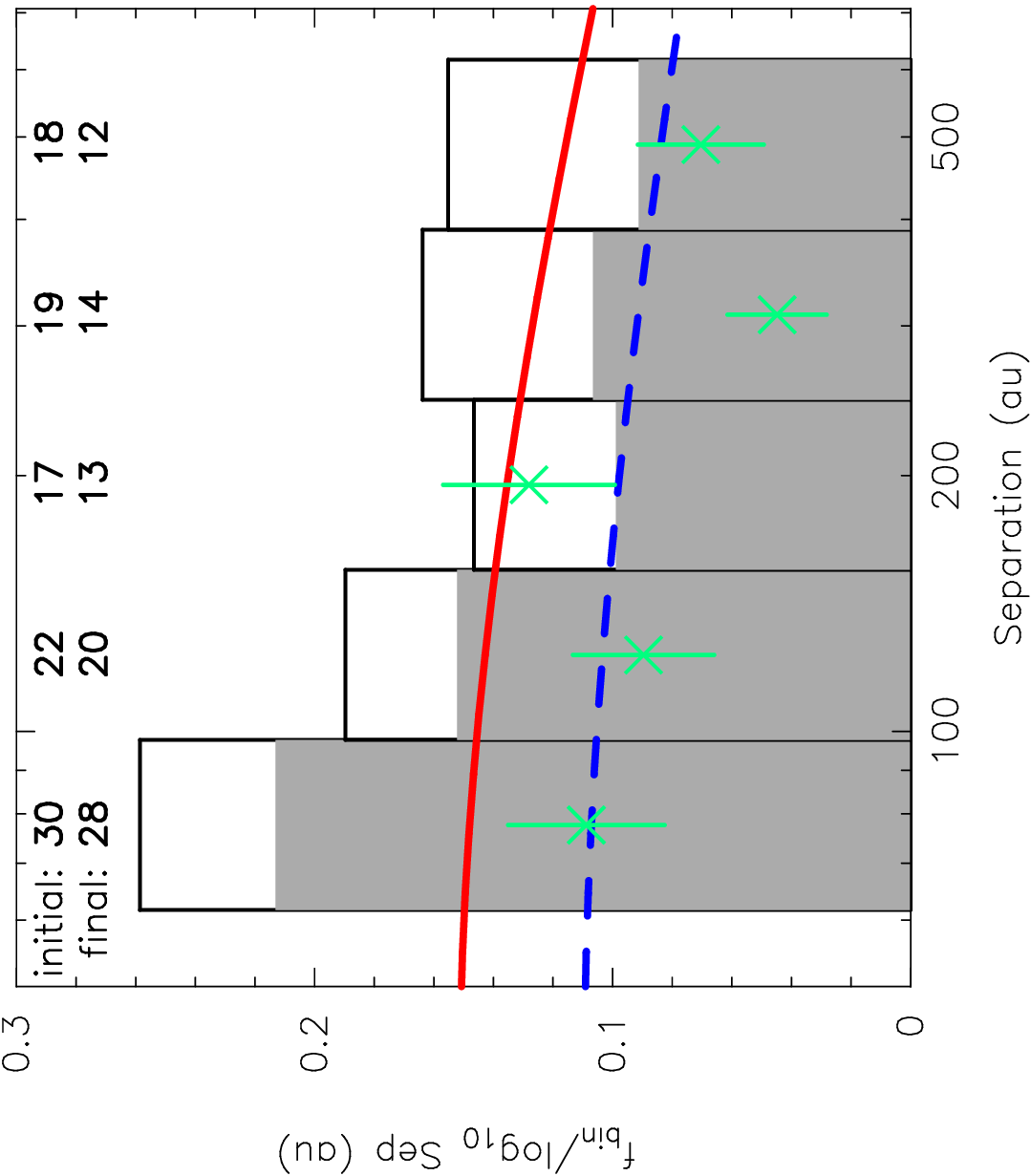}}}
\hspace*{0.3cm} 
\subfigure[]{\label{rei_sep-b}\rotatebox{270}{\includegraphics[scale=0.26]{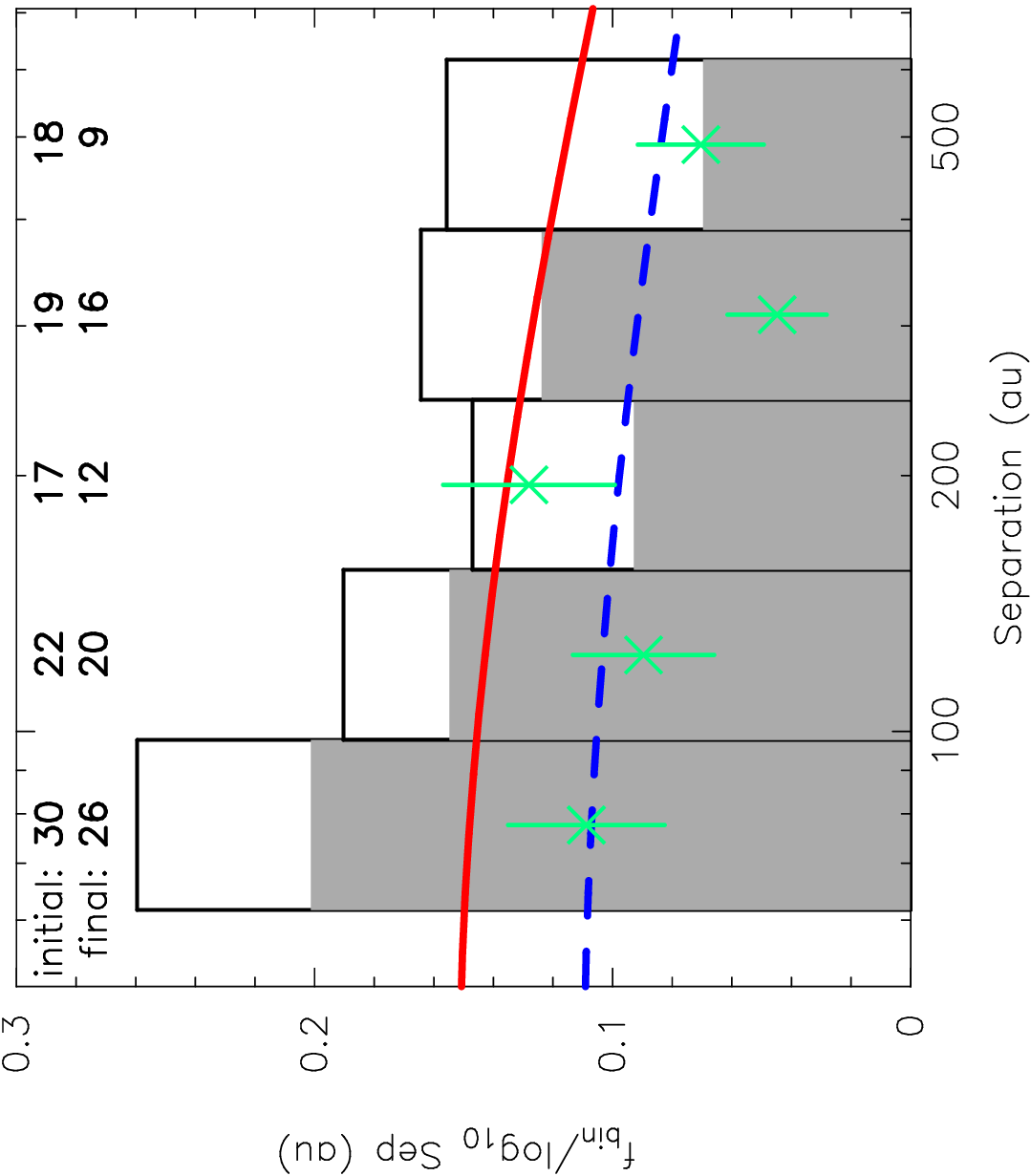}}}
\hspace*{0.3cm} 
\subfigure[]{\label{rei_sep-c}\rotatebox{270}{\includegraphics[scale=0.26]{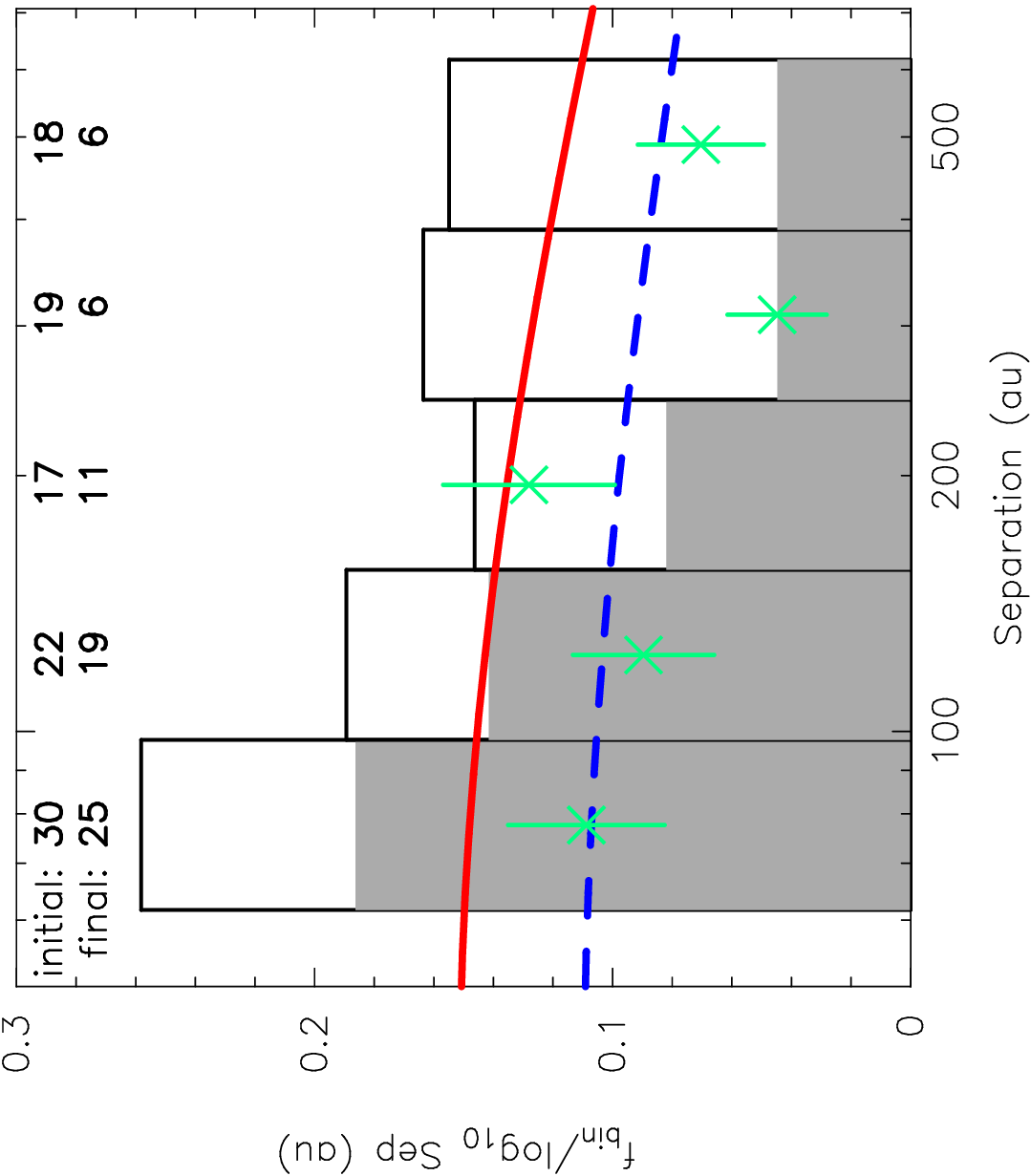}}}
\vspace*{-0.2cm}
\hspace*{-0.5cm}\subfigure[]{\label{rei_sep-d}\rotatebox{270}{\includegraphics[scale=0.26]{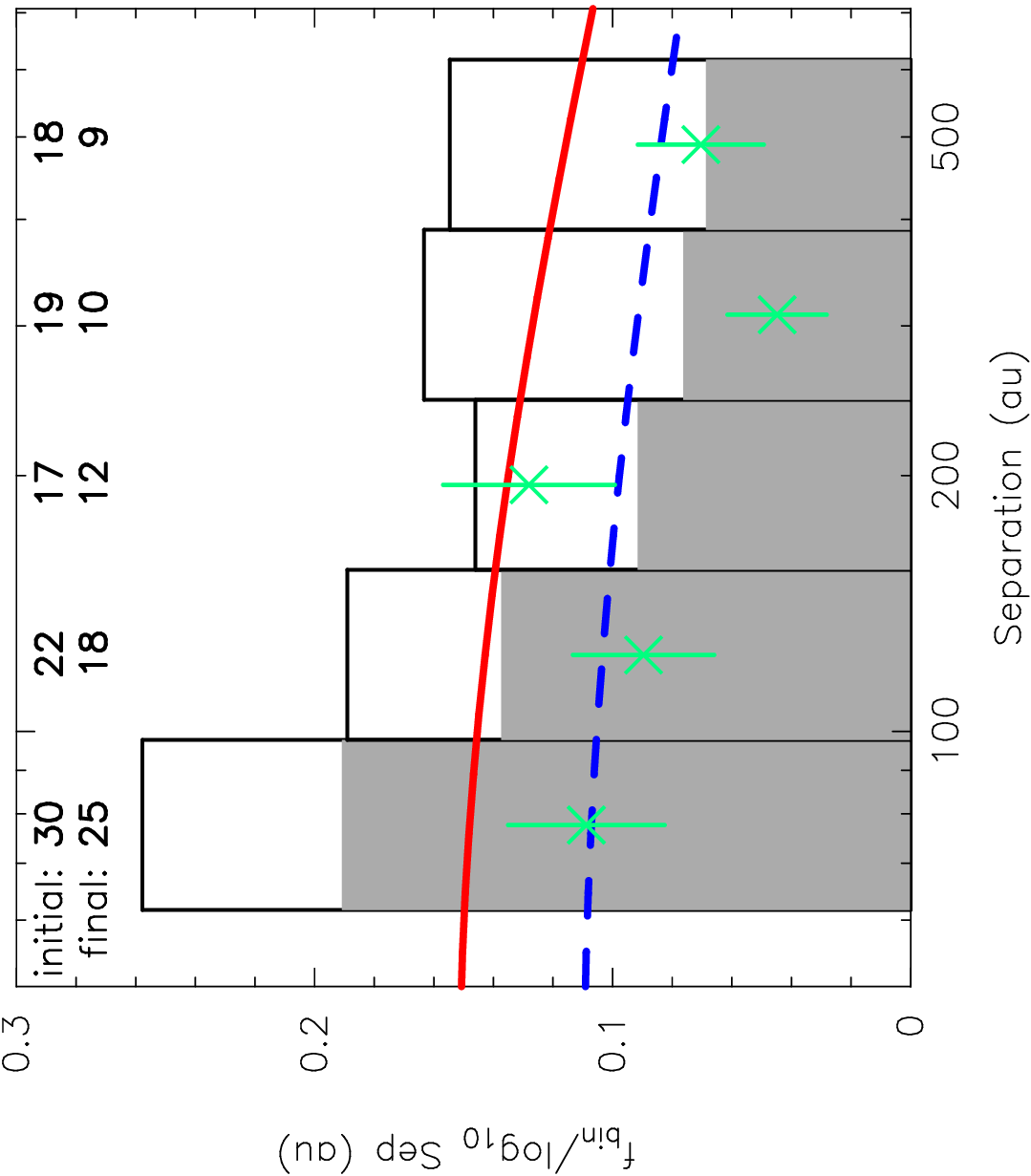}}}
\hspace*{0.3cm} 
\subfigure[]{\label{rei_sep-e}\rotatebox{270}{\includegraphics[scale=0.26]{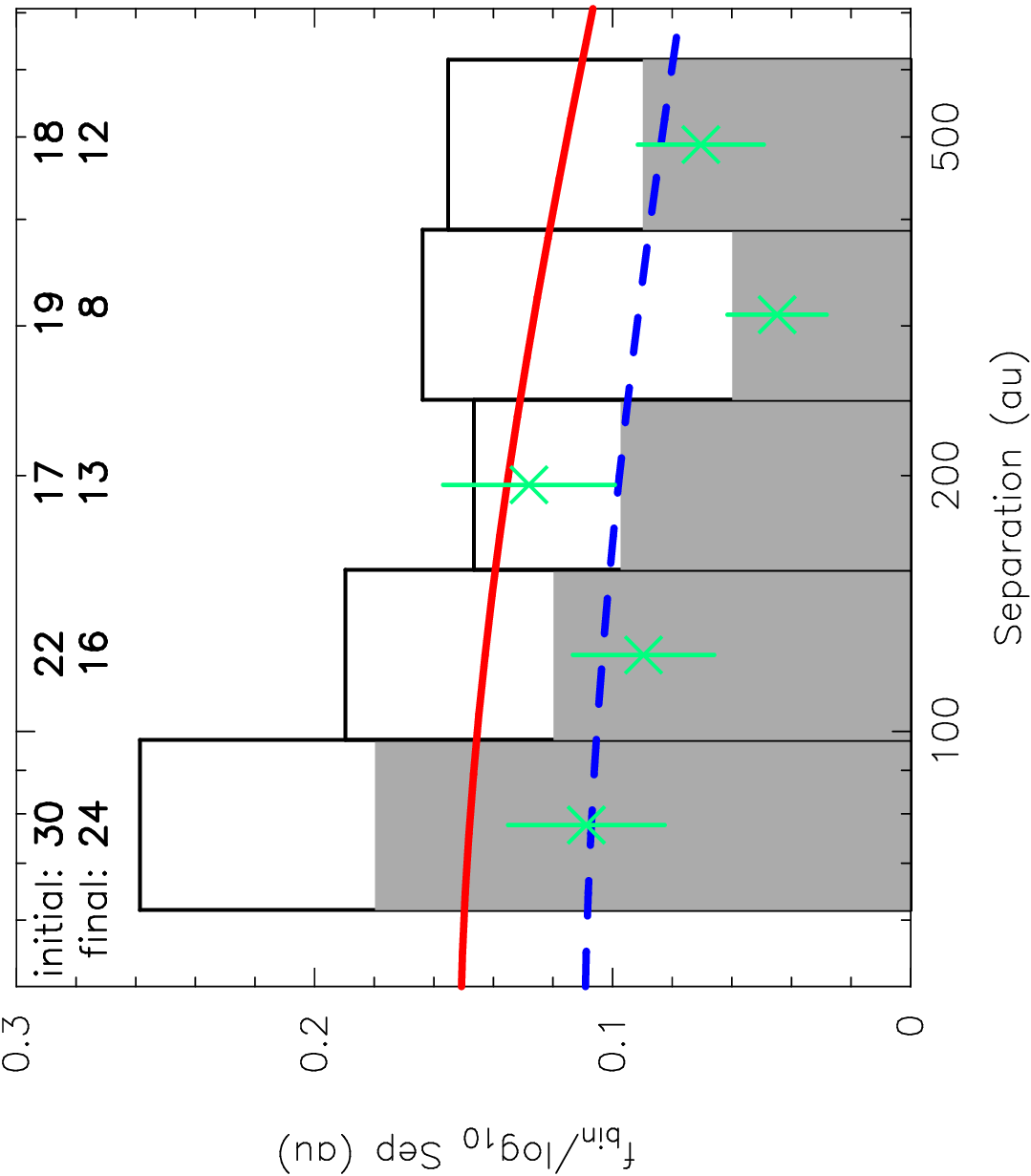}}}
\hspace*{0.3cm} 
\subfigure[]{\label{rei_sep-f}\rotatebox{270}{\includegraphics[scale=0.26]{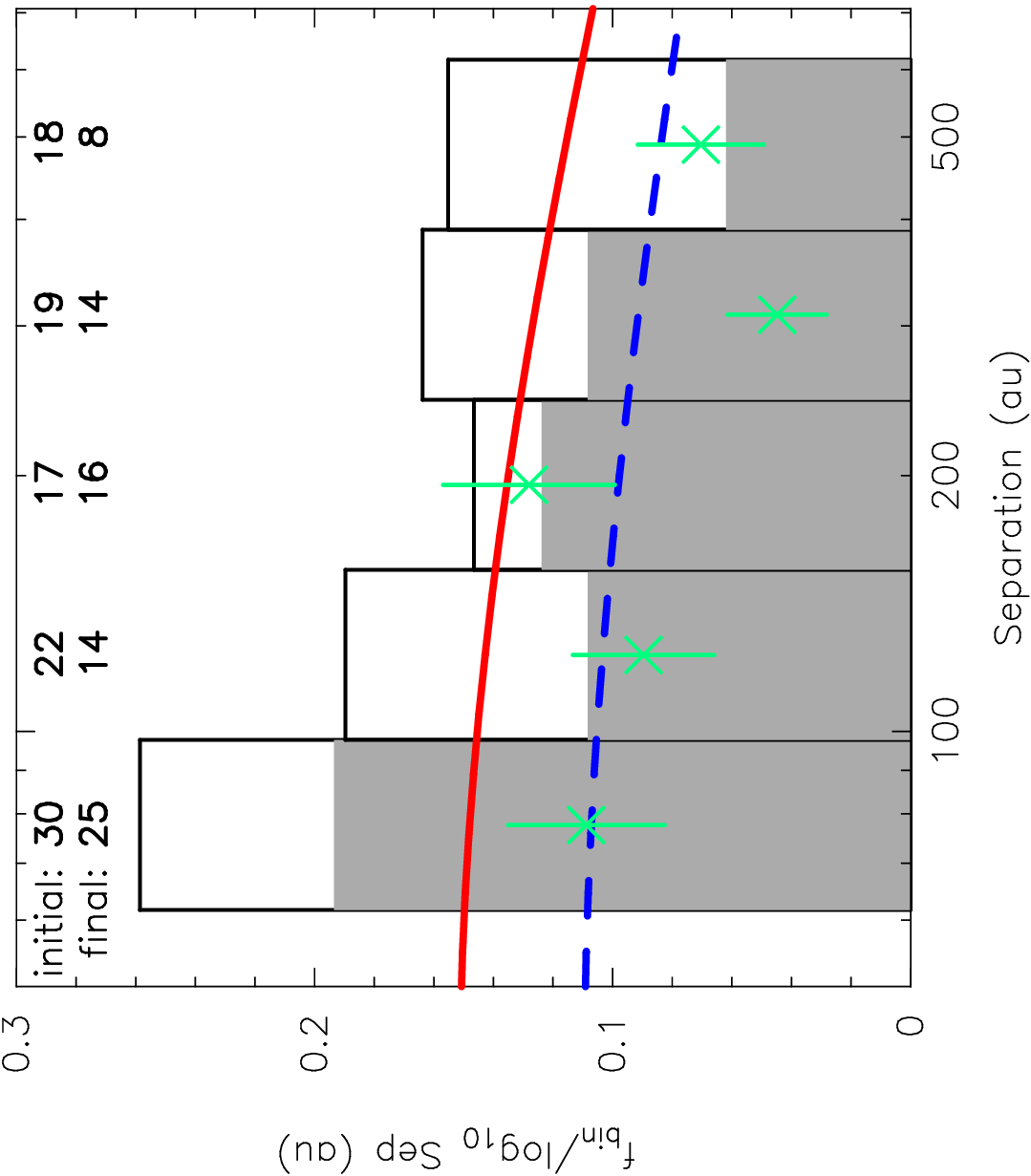}}}
\vspace*{-0.2cm}
\hspace*{-0.5cm}\subfigure[]{\label{rei_sep-g}\rotatebox{270}{\includegraphics[scale=0.26]{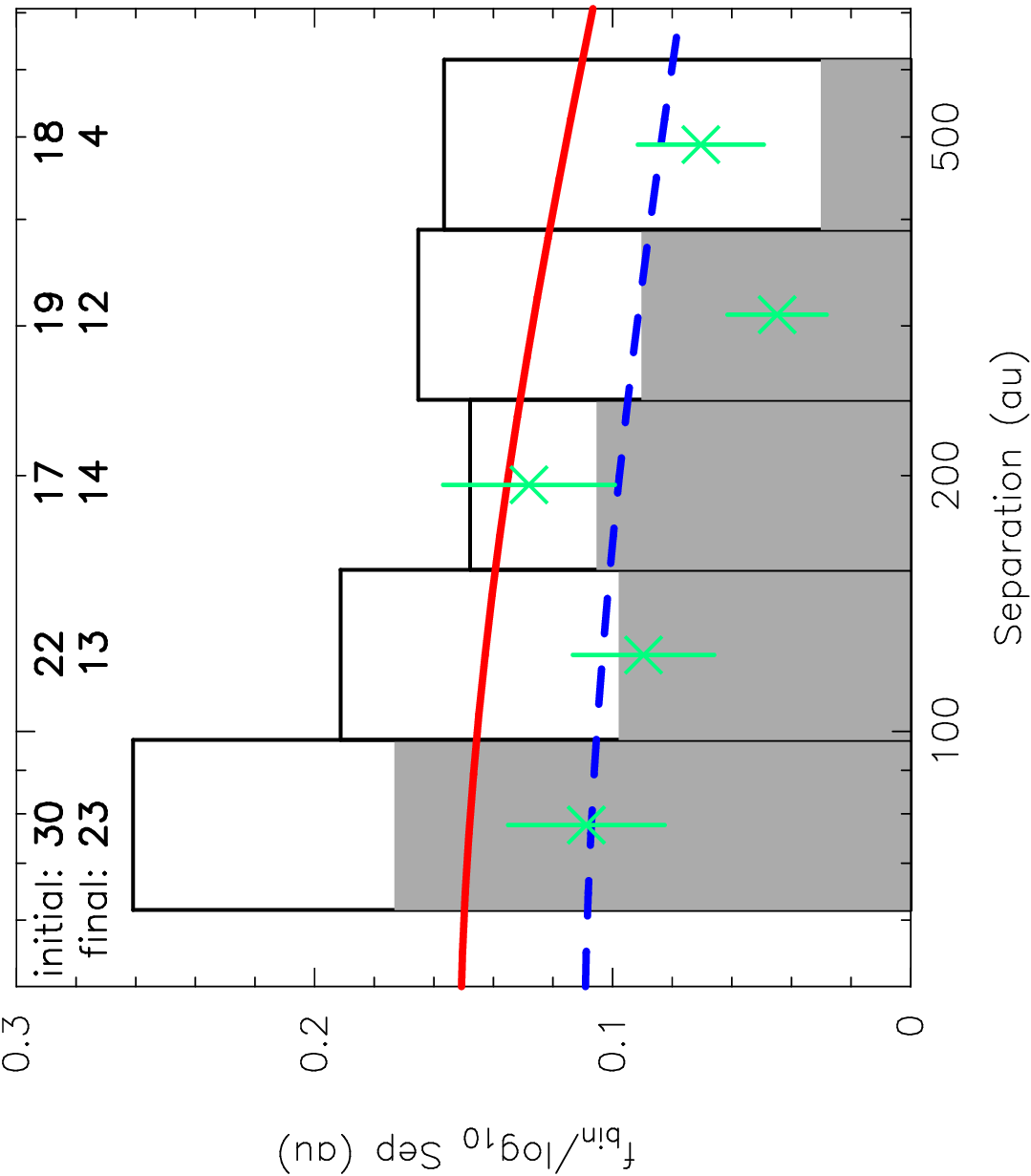}}}
\hspace*{0.3cm} 
\subfigure[]{\label{rei_sep-h}\rotatebox{270}{\includegraphics[scale=0.26]{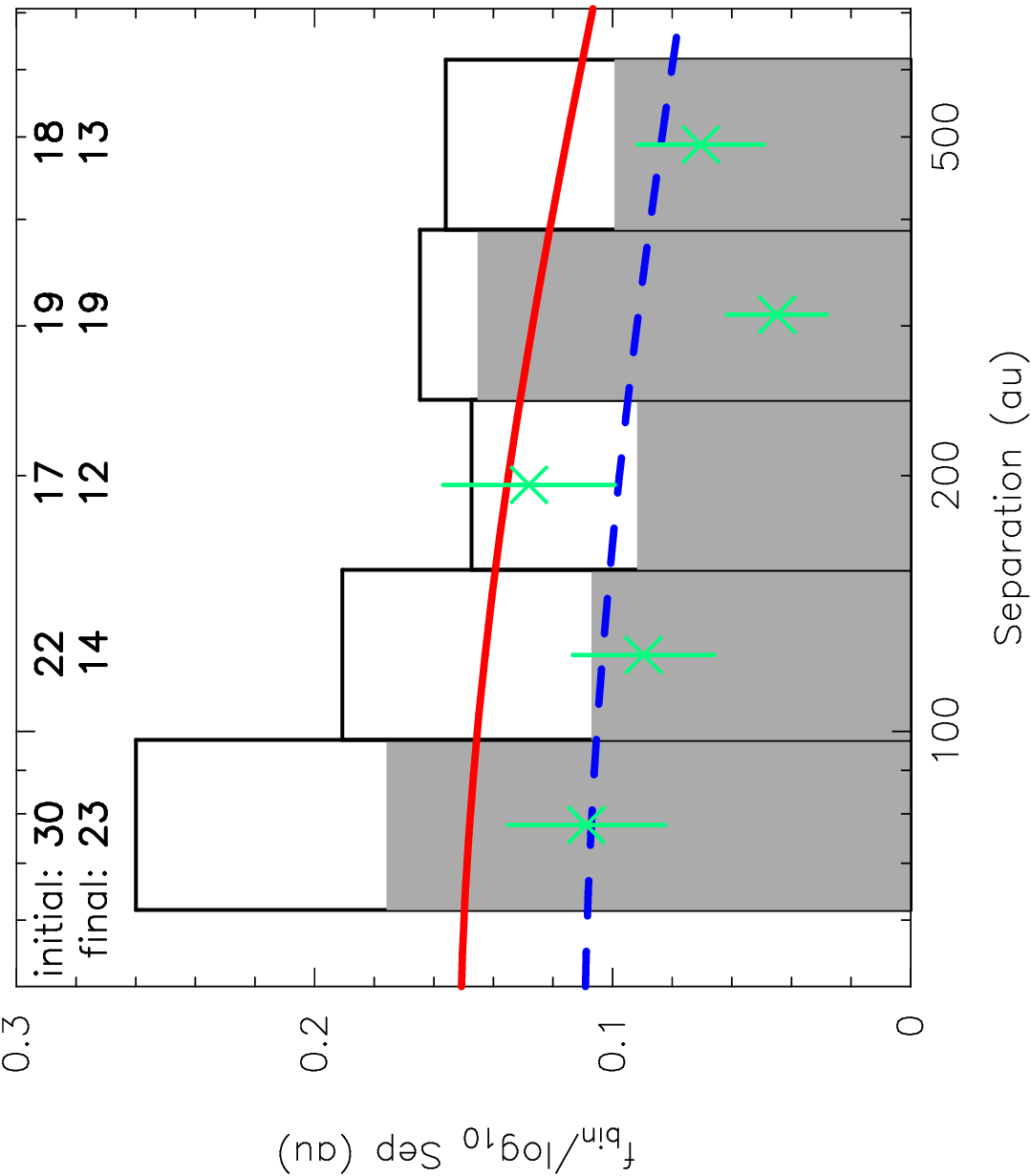}}}
\hspace*{0.3cm} 
\subfigure[]{\label{rei_sep-i}\rotatebox{270}{\includegraphics[scale=0.26]{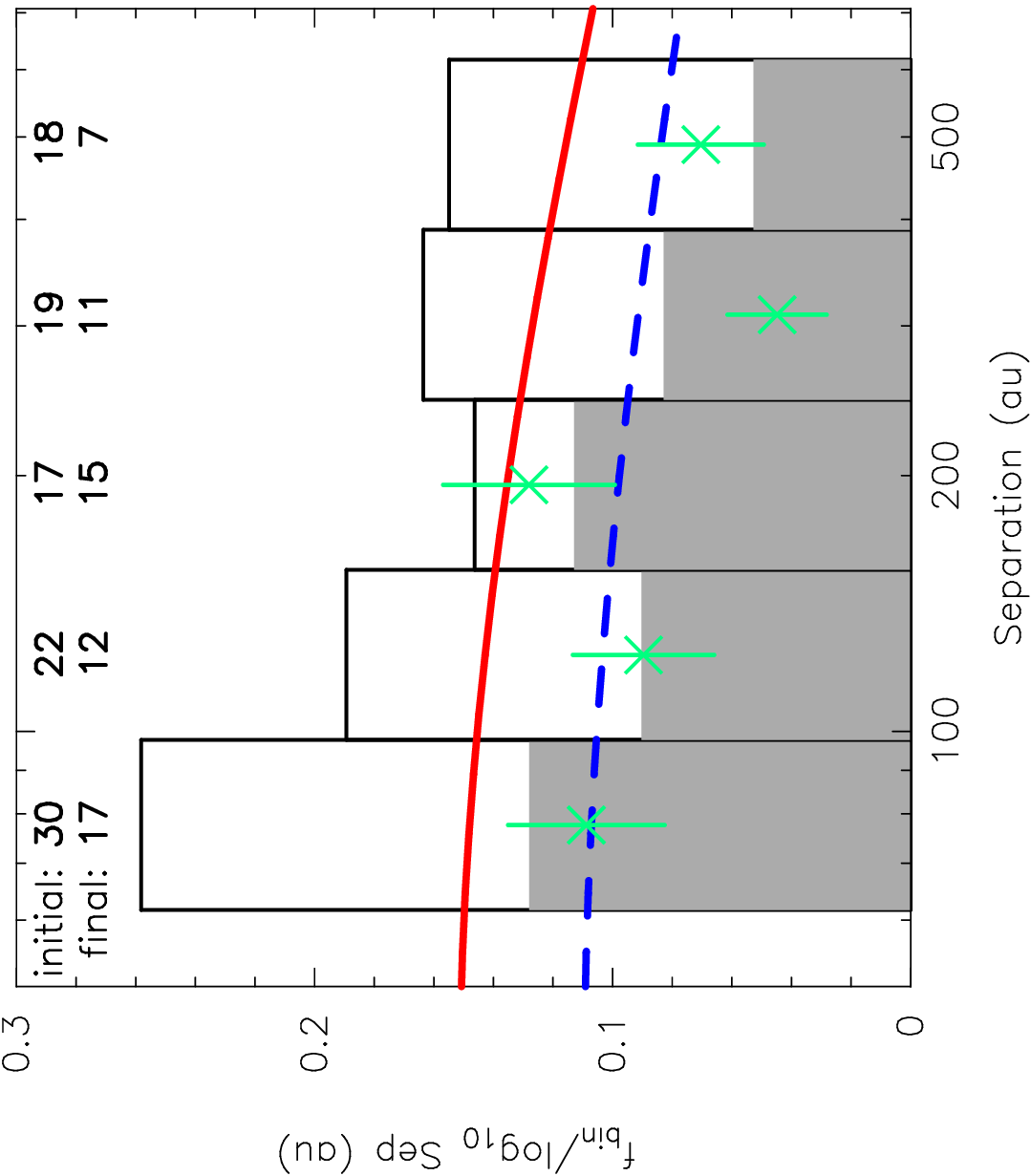}}}
\hspace*{-0.5cm} 
\subfigure[]{\label{rei_sep-j}\rotatebox{270}{\includegraphics[scale=0.26]{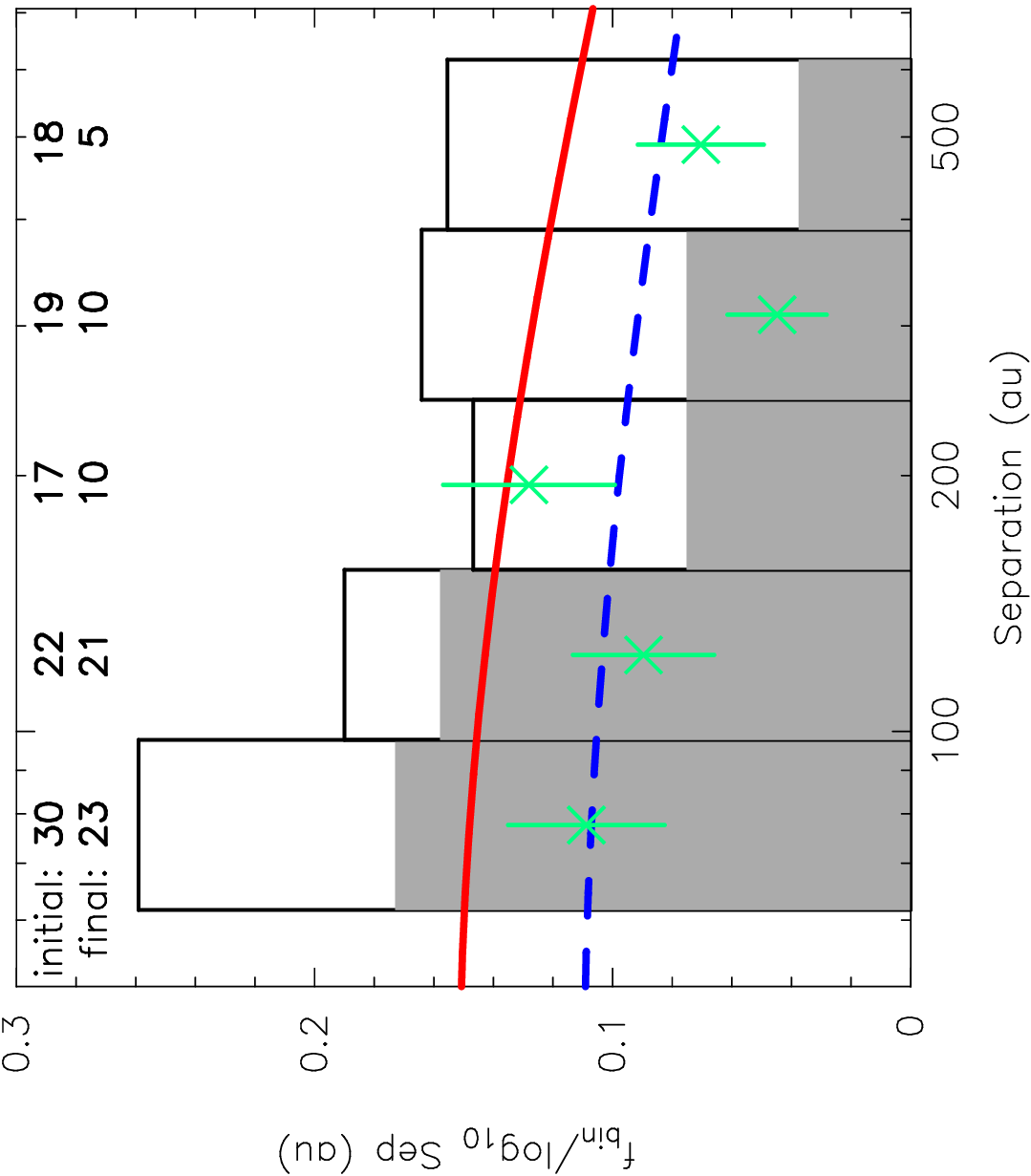}}}
\hspace*{0.3cm} 
\vspace*{-0.2cm}
\subfigure[AVERAGE]{\label{rei_sep-k}\rotatebox{270}{\includegraphics[scale=0.26]{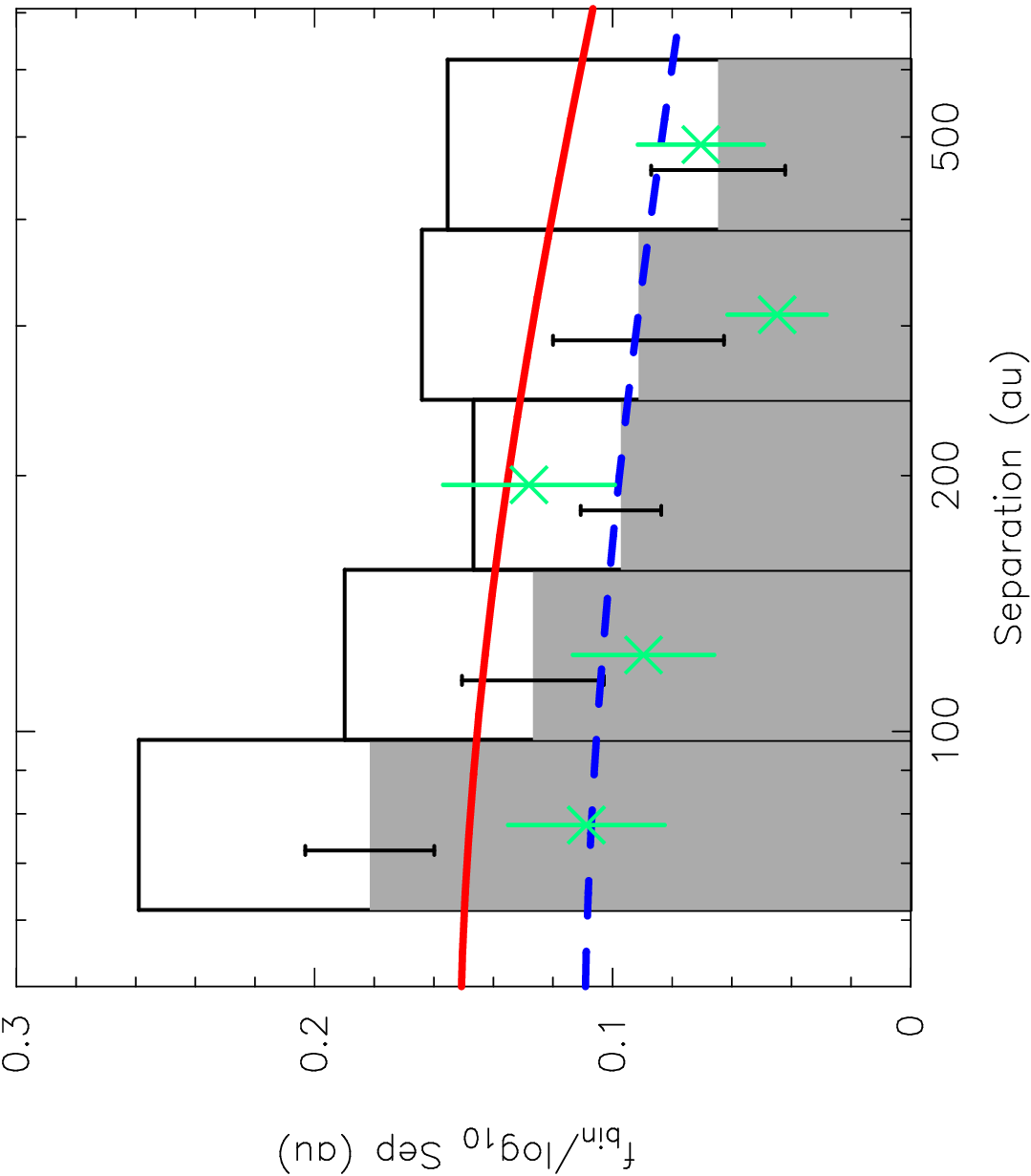}}}
\caption[bf]{Individual intermediate separation distributions in the
  range (62 -- 620\,au) probed by \citet{Reipurth07} (panels (a) --
  (j)).  The average of all 10 simulations is shown in panel (k). The
  (constant)  initial distribution is shown by the open histogram, and the final
  distribution is shown by the shaded histogram. The observations by
  \citet{Reipurth07} are shown by the green crosses. The log-normal
  fits to the field separation distributions  for G- and M-dwarfs are
  shown by the solid red and dashed blue lines, respectively.  Along the top are
  the initial and final numbers of binaries in each bin; note that the
initial numbers of binaries are always the same.}
\label{reipurth_dist}
  \end{center}
\end{figure*}

\begin{figure*}
  \begin{center}
\setlength{\subfigcapskip}{10pt}
\hspace*{-1.5cm}\subfigure[]{\label{rei_frac_cumul-a}\rotatebox{270}{\includegraphics[scale=0.35]{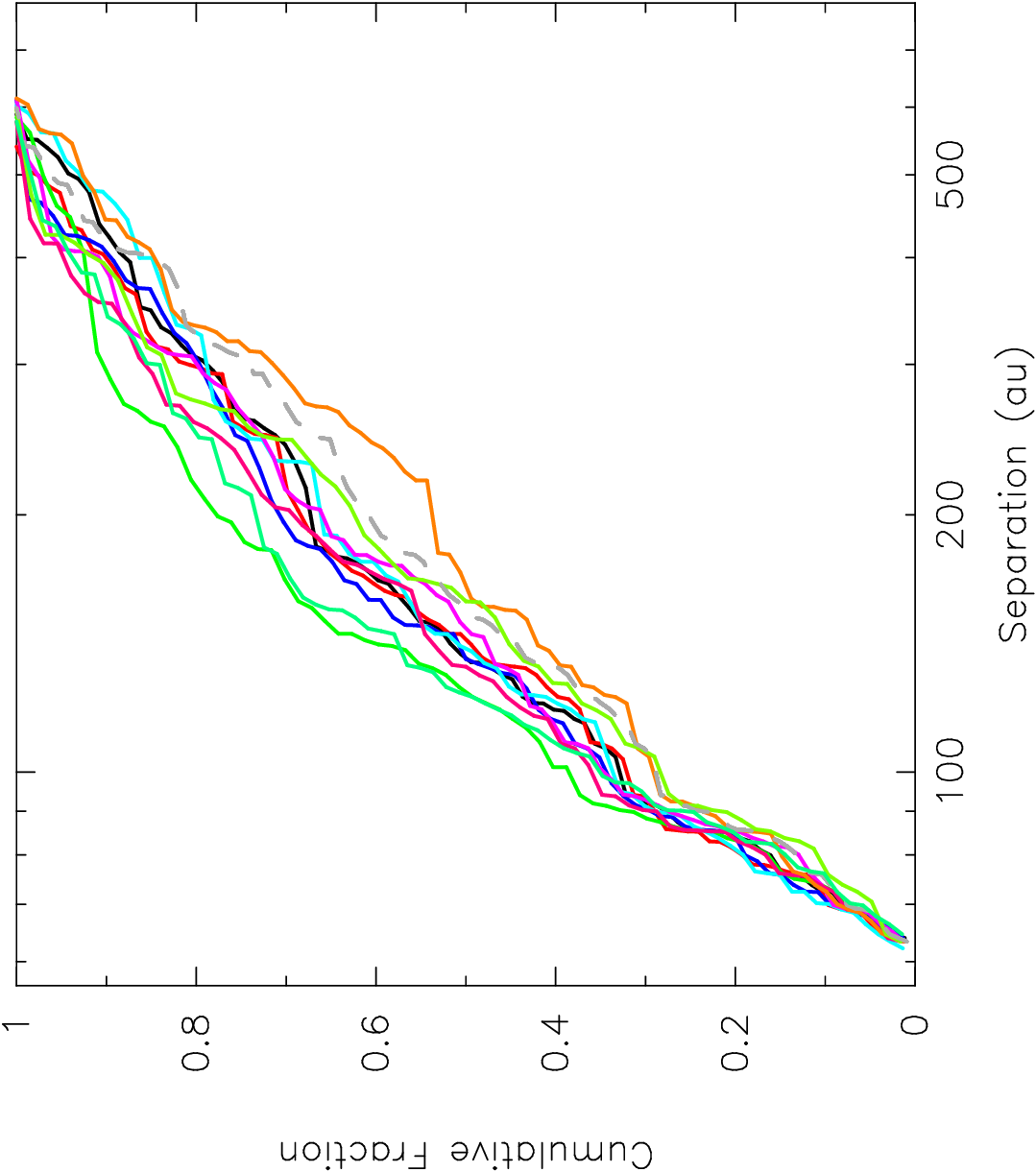}}}
\hspace*{0.3cm}
\subfigure[]{\label{rei_frac_cumul-b}\rotatebox{270}{\includegraphics[scale=0.35]{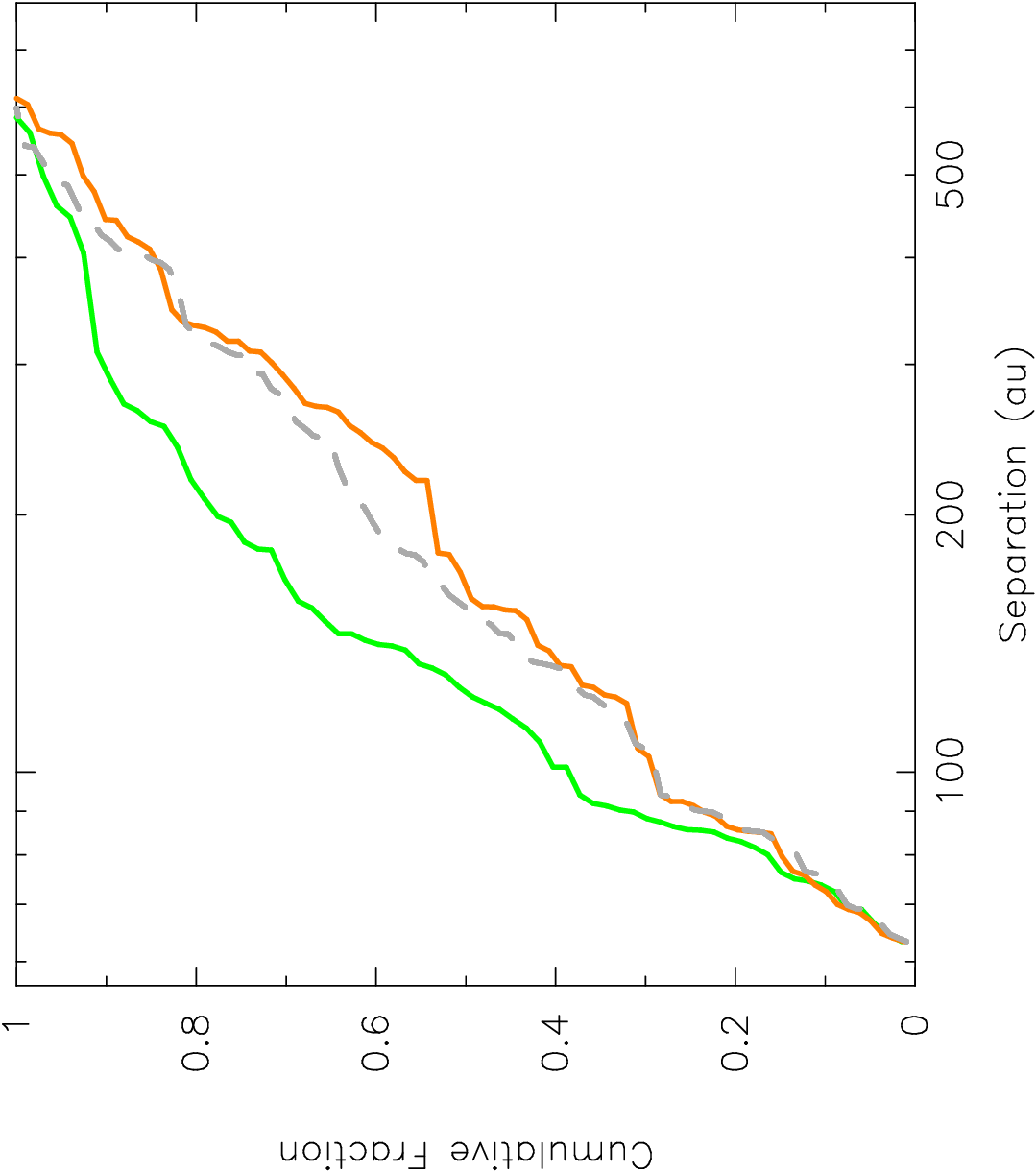}}}
\caption[bf]{The cumulative separation distribution of binaries in the
  separation range 62 -- 620\,au in (a) 10 different fractal clusters,
  and (b) the two extrema,  after 1\,Myr. The initial binary
  population is shown by the thick dashed grey line in both panels and
  is identical for each cluster.}
\label{reipurth_fractal_cumulative}
  \end{center}
\end{figure*}

However,  in both morphologies the final binary fractions can differ by almost a factor of two, irrespective of density.

\subsection{Separation distribution}

As well as changing the binary fraction in the intermediate 62 --
620\,au range, the distribution of separations can also be changed
significantly.

In Fig.~\ref{reipurth_dist} we show the individual separation 
distributions in the range 62 -- 620\,au for each fractal cluster, 
binned in the same way as the data in
\citet{Reipurth07}.  The initial separation  distribution, which is identical
in each simulation, is shown by the open histogram, and the
separation distribution after 1\,Myr is shown by the shaded 
histogram. For comparision,  the data from \citet{Reipurth07} is shown
by the green crosses, and the log-normal fits to the field separation
distributions for G- and M-dwarfs are shown by the solid red and
dashed blue  lines, respectively. The average of all 10 simulations,
with 1--$\sigma$ uncertainties, is shown in panel (k). 

\begin{figure}
 \begin{center}
\rotatebox{270}{\includegraphics[scale=0.39]{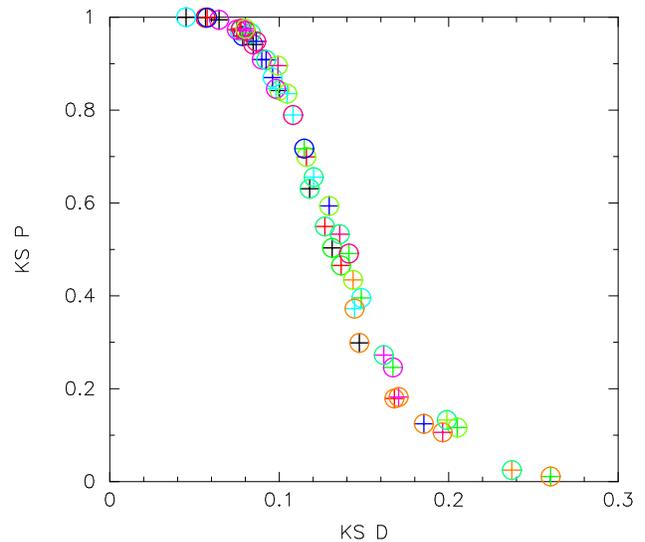}}
\caption[bf]{The distribution of values for KS tests between all
  pairs of the fractal cluster simulations on the cumulative separation distributions (colours correspond to those in Fig.~\ref{reipurth_fractal_cumulative}). We show the KS p-value against the KS $\mathcal{D}$ statistic.}
\label{fractal_KS_seps}
\end{center}
\end{figure}

As noted in \citet{Parker11c}, averaging together the 10 realisations
of clusters with a fractal morphology and subvirial velocities reproduces the
observed ONC binary distribution reasonably well \citep[see also][]{King12a}. However, from inspection of Fig.~\ref{reipurth_dist}
we see that the same initial population can evolve to very different
distributions over the course of 1\,Myr.  Clearly, the initial 
population of 106 intermediate binaries is processed
differently in each cluster.

In Fig.~\ref{reipurth_fractal_cumulative} we show the cumulative
distributions of the intermediate binary separations along with 
the initial separation
distribution (the thick dashed grey line in both panels).  In
Fig.~\ref{rei_frac_cumul-a} all ten fractal realisations
are shown, in Fig.~\ref{rei_frac_cumul-b} we show the
two most different cumulative distributions.

In Fig.~\ref{fractal_KS_seps} we plot the 45 possible KS-test comparisons between the cumulative separation distributions of the 
fractal clusters. We reject the null hypothesis of there being no difference between two separation distributions if the KS 
p-value is less than 0.05.

The two most different separation distributions are from the
distributions shown in panels (c) and (h) of
Fig.~\ref{reipurth_dist}.  Note that these two simulations are not the
same two simulations that produce the largest difference in binary
fraction (which were those in panels (a) and (i)).

\begin{figure*}
  \begin{center}
\setlength{\subfigcapskip}{10pt}
\hspace*{-1.5cm}\subfigure[]{\label{rei_plummer_cumul-a}\rotatebox{270}{\includegraphics[scale=0.35]{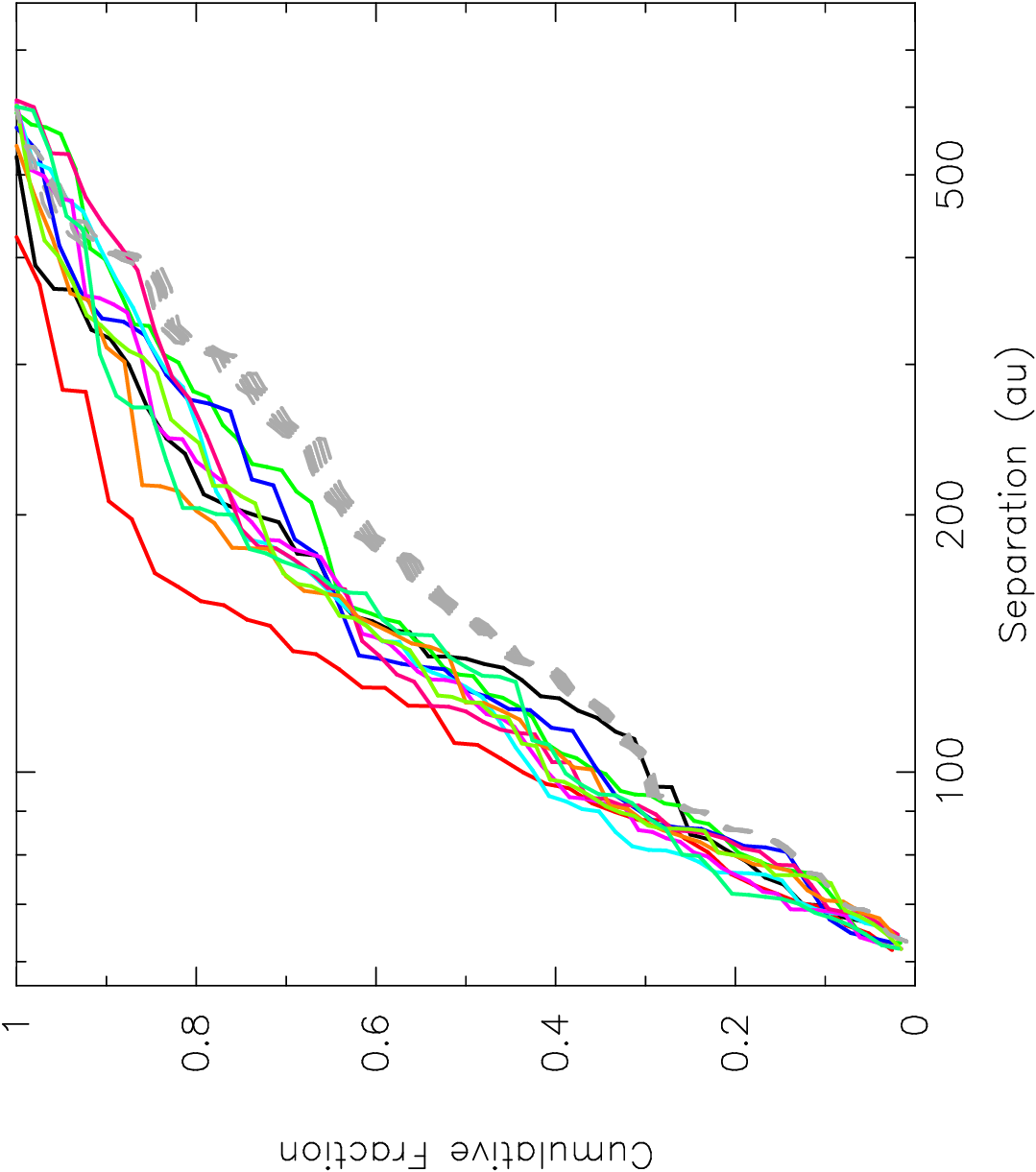}}}
\hspace*{0.3cm}
\subfigure[]{\label{rei_plummer_cumul-b}\rotatebox{270}{\includegraphics[scale=0.35]{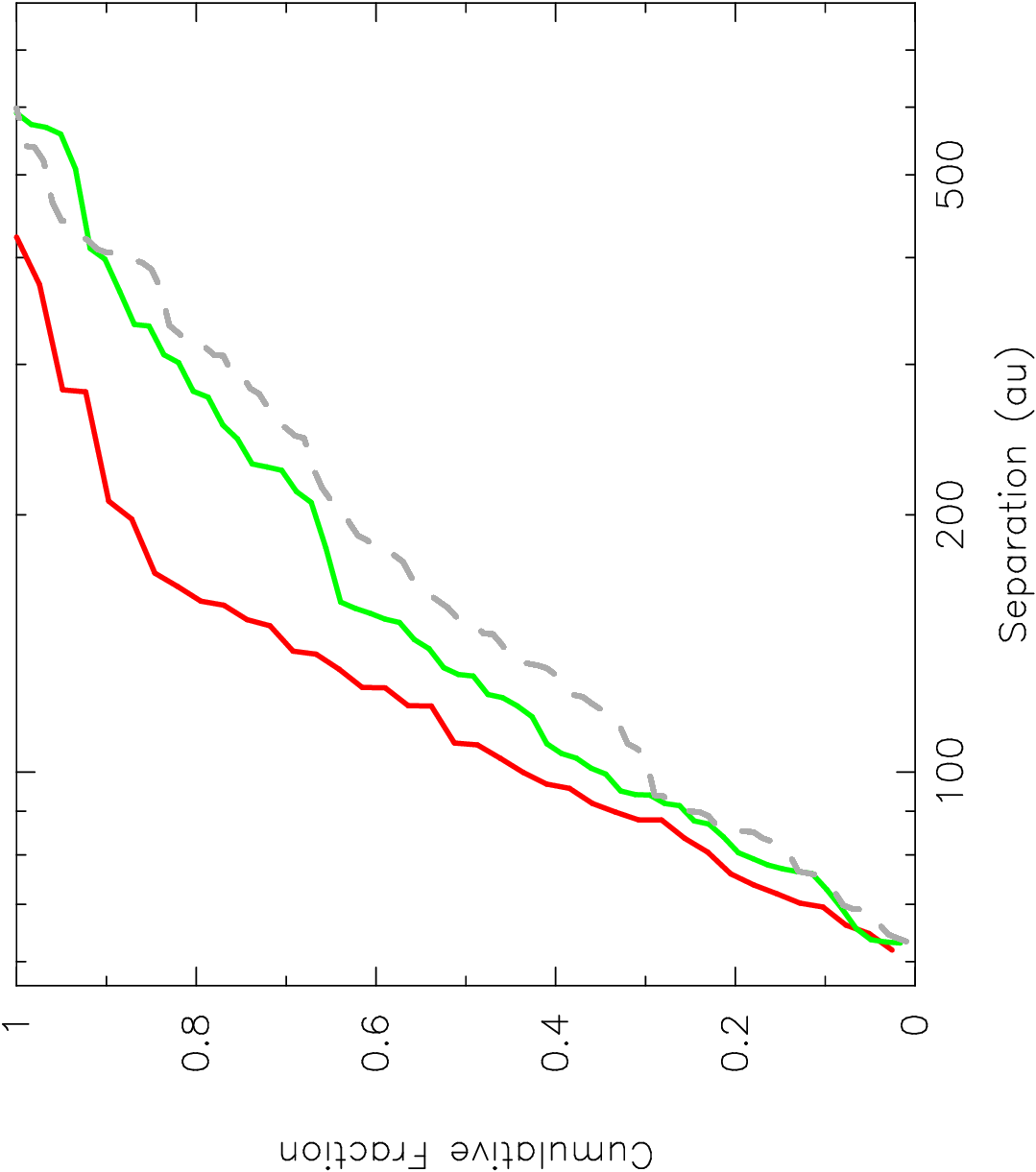}}}
\caption[bf]{The cumulative separation distribution of binaries in the
  separation range 62 -- 620\,au in (a) 10 different Plummer-sphere
  clusters, and (b) the two extrema,  after 1\,Myr.  In panel (a) the
  thick dashed grey lines show the initial distributions which are
  slightly different for each cluster (see text), the differences are
  so small that in panel (b) we only plot the primordial distribution
 for the righthand cluster for clarity.}
\label{reipurth_Plummer_cumulative}
  \end{center}
\end{figure*}

The largest difference between the two extreme distributions is
$\mathcal{D}=0.26$; for this $\mathcal{D}$ a KS test gives the p--value $\mathcal{P}=0.01$.  
This is a very significant difference and one would draw
the (correct) conclusion that these two distributions are different.
However, it is not the {\em initial} distributions that were different
(they were identical), rather it is the dynamical processing of the
systems that was very different.

\begin{figure}
 \begin{center}
\rotatebox{270}{\includegraphics[scale=0.39]{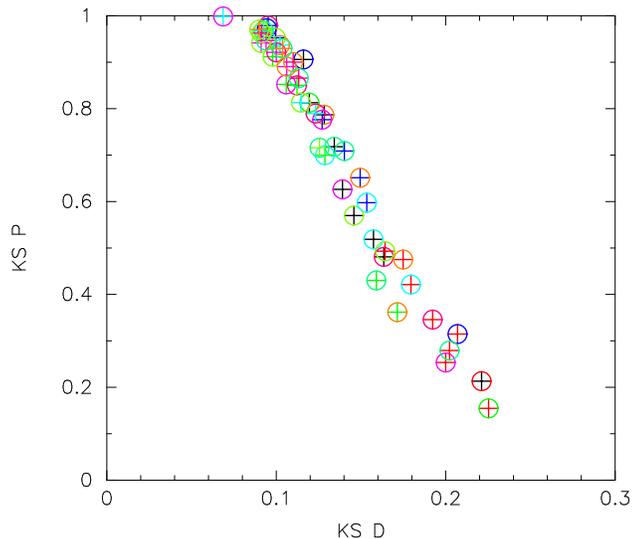}}
\caption[bf]{The distribution of values for KS tests between all
  pairs of the Plummer-sphere cluster simulations on the cumulative separation distributions (colours correspond to those in Fig.~\ref{reipurth_Plummer_cumulative}). 
We show the KS p-value against the KS $\mathcal{D}$ statistic.}
\label{Plummer_KS_seps}
\end{center}
\end{figure}

Note that only two pairs of simulations are rejected by the KS test as
being significantly different (pairs c and h, and h and j), with
simulation h being the most significant outlier.  
That nine simulations agree reasonably well with each other (for the
separation distribution, not necessarily, as we have seen, for binary
fractions) shows that this simulation is just an outlier.  
However, the problem is that in just ten simulations we have produced a significant outlier, and there is no way of telling if a single observed distribution for a cluster is such an outlier.

Examination of Fig.~\ref{rei_frac_cumul-b} shows that
one extreme (the lower orange line) remains very close to the initial
separation distribution (the thick dashed grey line).  This
distribution is from panel (h) of Fig.~\ref{reipurth_dist} and it can
be seen that it has lost roughly the same fraction of binaries from
each bin so retaining the shape of the initial distribution.  The
upper extreme (the green line) shows a very different separation
distribution to the initial distribution, as can also be seen in panel (c)
of Fig.~\ref{reipurth_dist} this cluster has mainly lost wider
binaries ($> 200$~au).

Generally speaking (and visible in Figs.~\ref{reipurth_dist}
and~\ref{reipurth_fractal_cumulative}), wider binaries are more
susceptable to disruption as they are more weakly bound.  A key result
however is that because binary destruction is stochastic, the probability of
destruction of an {\em intermediate} system depends more on if a 
system has had a close encounter or not than on the binding energy of
the system \citep[see][]{Hills75a,Heggie75}.

One could hypothesise that the very different and stochastic dynamical
histories of different fractals \citep[see][]{Allison10} might 
be responsible for the very large differences
in the resulting populations.  To test this in 
Fig.~\ref{reipurth_Plummer_cumulative} we plot the cumulative
distributions of separations in our ten Plummer spheres for 
all ten realisations (panel (a)) and the two extremes (panel (b))
after 1~Myr along with the initial distribution (the thick
dashed grey lines).  Note that the initial distributions for each
Plummer sphere are very slightly different, this is because some
intermediate systems are in such a place that they are not identified
as binaries even at time zero.  Different 
realisations of Plummer spheres are almost
impossible to distinguish, and their dynamical histories will be very
similar.  The only differences should be in the chance of a particular
system having (a) destructive encounter(s) or not.

Again we find a wide spread in the distributions,  and we plot the results of the KS tests between each 
set of simulations in Fig.~\ref{Plummer_KS_seps}. Interestingly, although the Plummer spheres are more dense, the most 
extreme separation distributions (shown in Fig.~\ref{rei_plummer_cumul-b}) have $\mathcal{D}=0.23$ and a KS test 
p--value of $\mathcal{P} = 0.15$. Based on this value, we cannot reject the hypothesis that there is no difference between these separation distributions.  However, as we have seen, the binary fractions are also significantly affected, and care must be taken not to take a marginal result for the KS test in separation distribution together with different binary fractions to make us suspect a real difference between clusters.

This may indicate that the differences in the final separation distributions are due to the stochastic nature of the substructured fractals. However, as young, unevolved clusters 
appear to be substructured \citep{Cartwright04,Sanchez09}, and this substructure can disrupt binaries \citep{Parker11c,King12a}, then we must recognise that this stochasticity 
could affect our interpretation of observations of real clusters.  

\subsection{Mass ratio distribution}

\begin{figure*}
  \begin{center}
\setlength{\subfigcapskip}{10pt}
\hspace*{-1.5cm}\subfigure[]{\label{qrat_fractal_cumul-a}\rotatebox{270}{\includegraphics[scale=0.35]{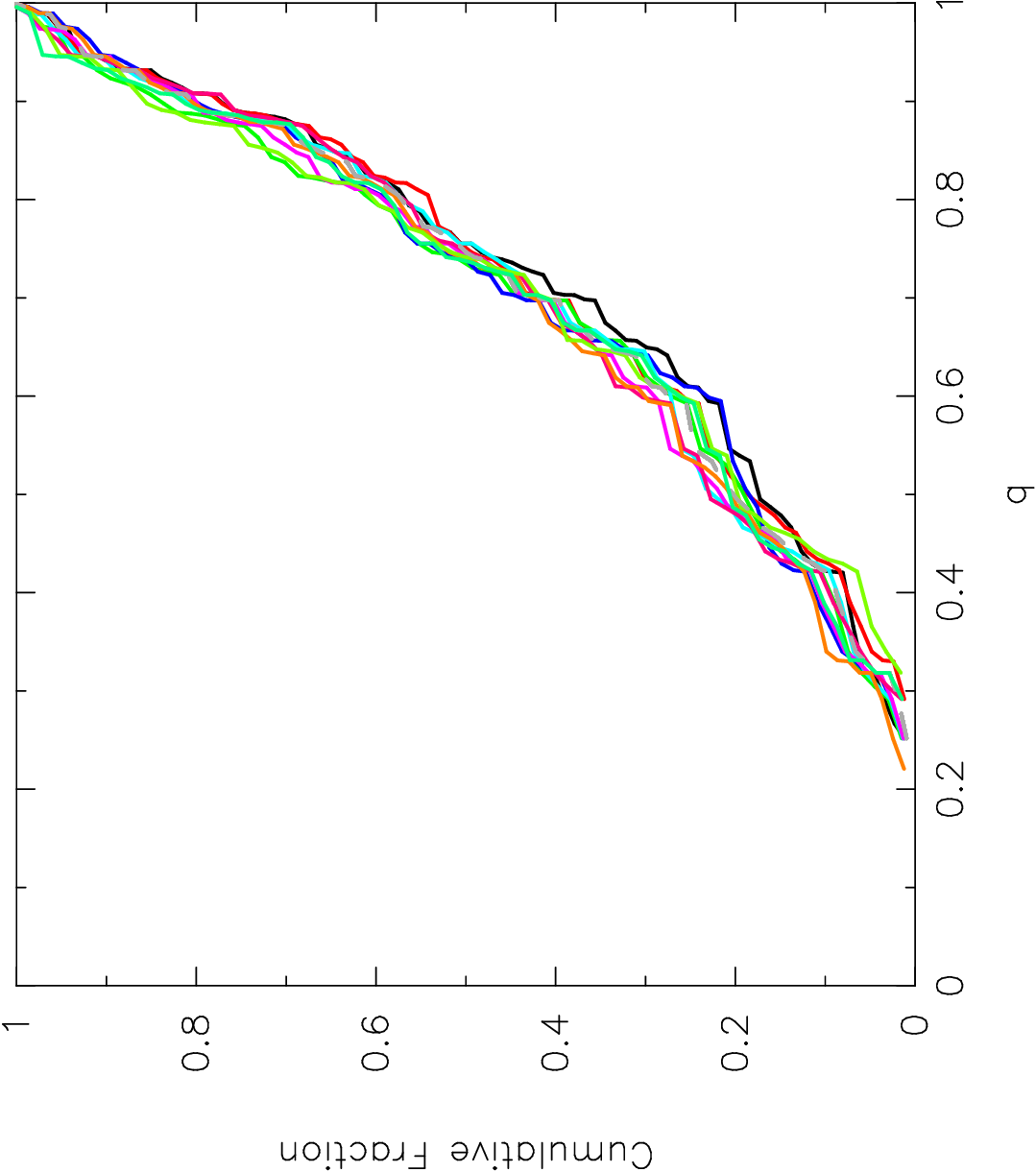}}}
\hspace*{0.3cm}
\subfigure[]{\label{qrat_fractal_cumul-b}\rotatebox{270}{\includegraphics[scale=0.35]{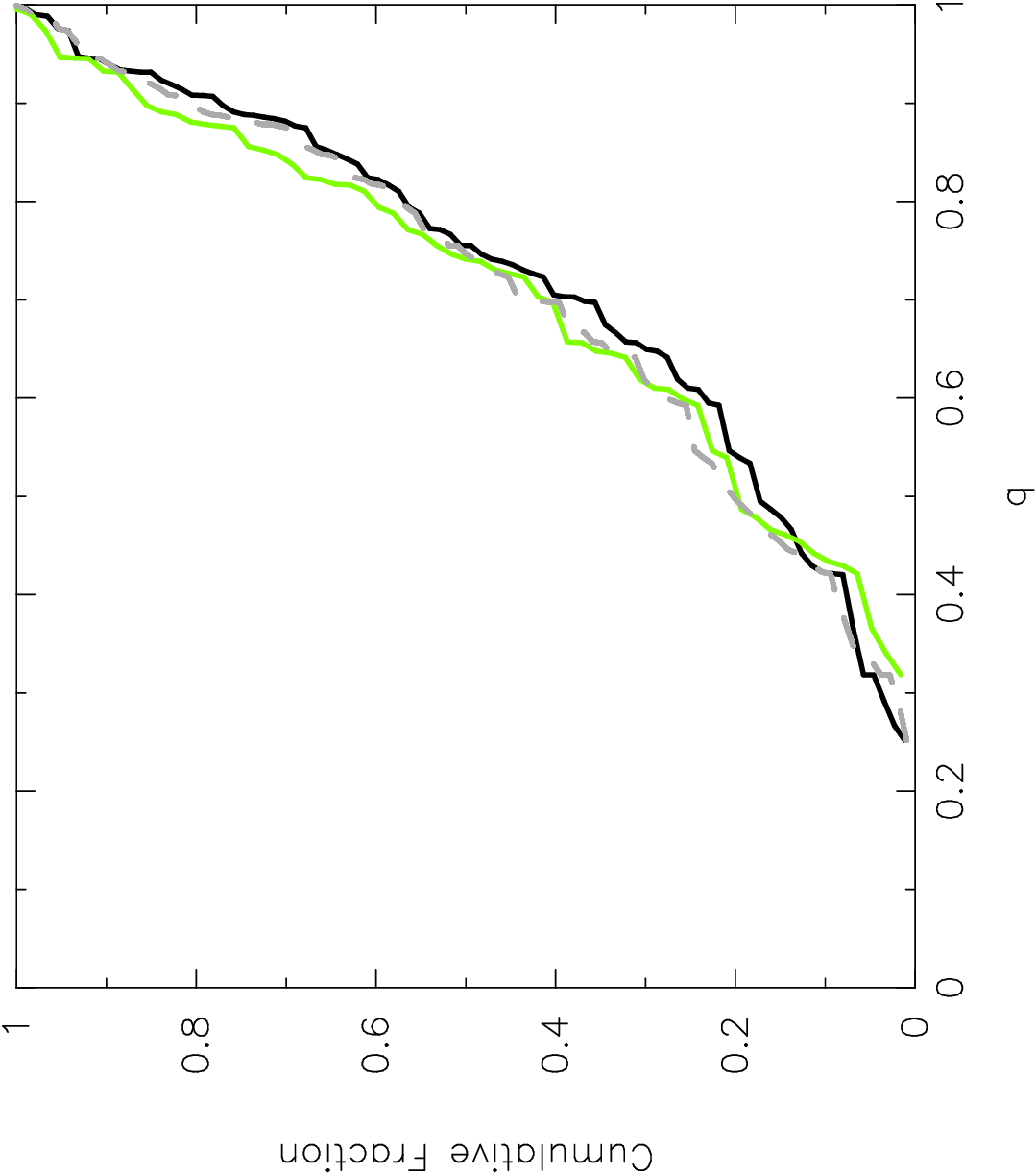}}}
\caption[bf]{The cumulative mass ratio distribution of binaries in the
  separation range 62 -- 620\,au in (a) 10 different fractal clusters,
  and (b) the two extrema, after 1\,Myr.  The initial distribution is
  shown by the thick dashed grey line in both panels.}
\label{qrat_fractal_cumulative}
  \end{center}
\end{figure*}

\begin{figure*}
  \begin{center}
\setlength{\subfigcapskip}{10pt}
\hspace*{-1.5cm}\subfigure[]{\label{qrat_Plummer_cumul-a}\rotatebox{270}{\includegraphics[scale=0.35]{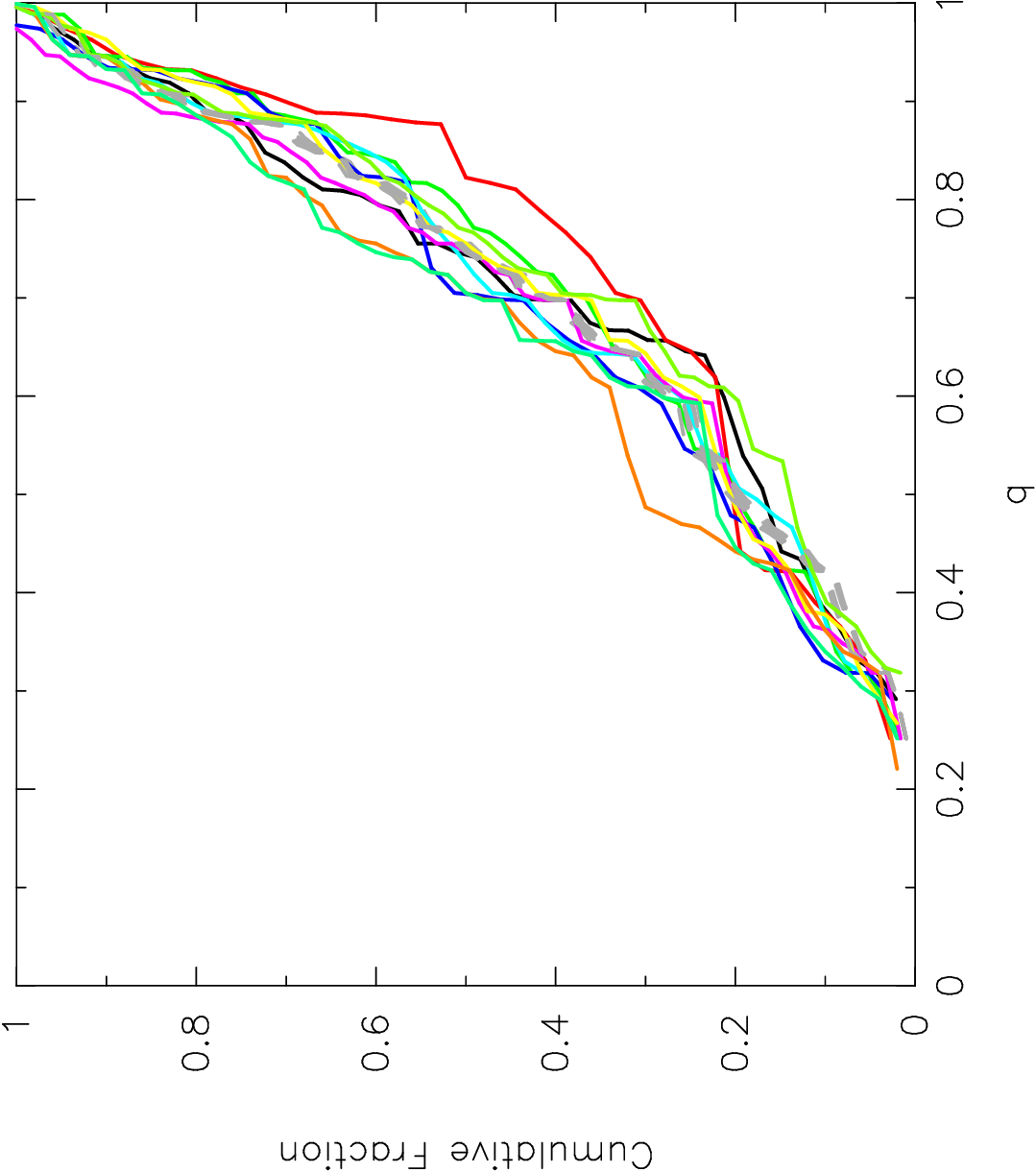}}}
\hspace*{0.3cm}
\subfigure[]{\label{qrat_Plummer_cumul-b}\rotatebox{270}{\includegraphics[scale=0.35]{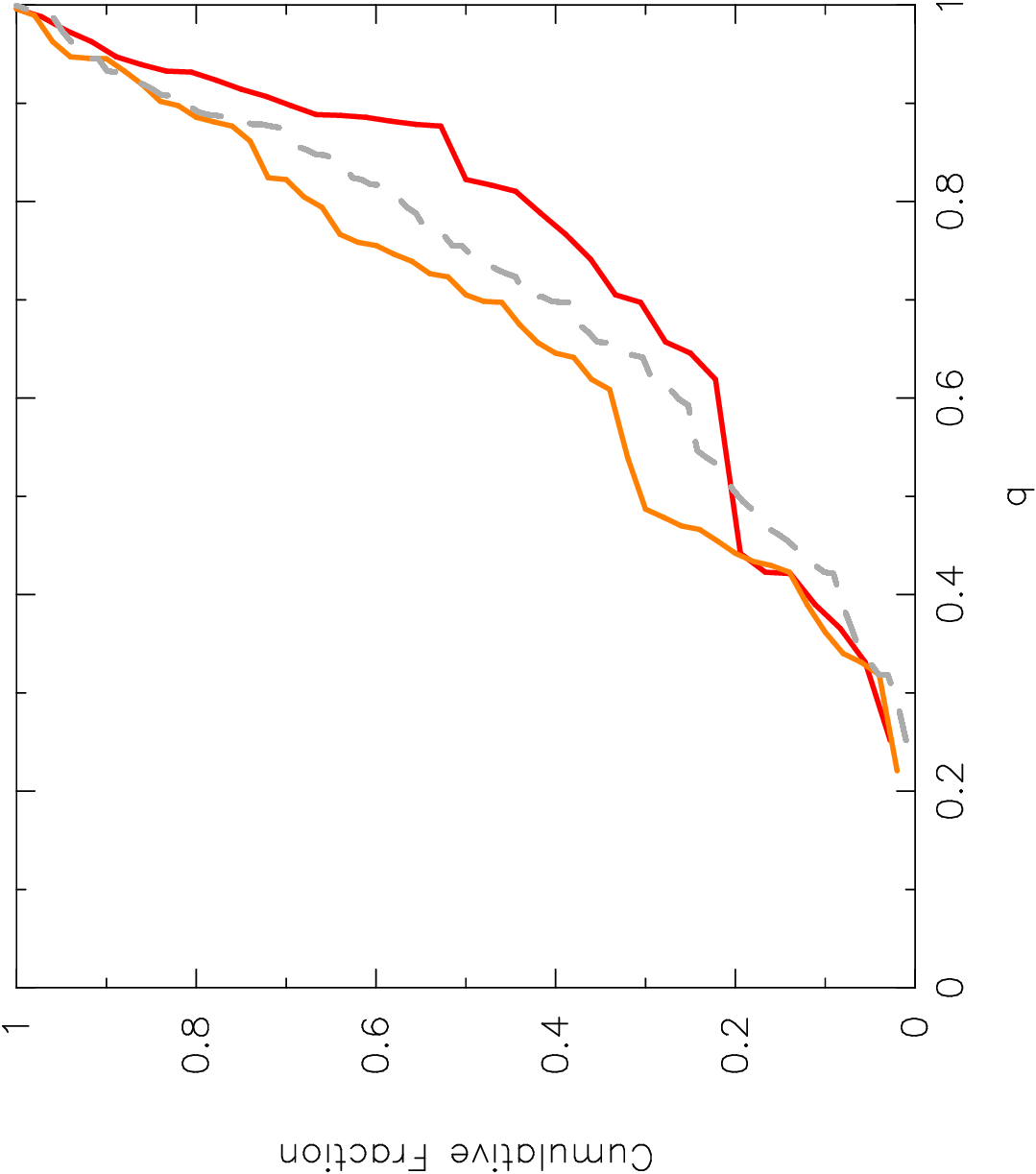}}}
\caption[bf]{The cumulative mass ratio distribution of binaries in the
  separation range 62 -- 620\,au in (a) 10 different Plummer-sphere
  clusters, and (b) the two extrema, after 1\,Myr. The initial
  distributions are shown by the thick dashed grey line in panel (a),
  but in panel (b)  the initial distribution for the righthand cluster
  only is shown for clarity. }
\label{qrat_Plummer_cumulative}
  \end{center}
\end{figure*}

We have seen that the numbers and separation distributions of
intermediate binaries can be significantly altered in a highly
stochastic way by encounters.  We now turn our attention to the mass
ratio distribution: is this also significantly altered, or does it
retain an imprint of the initial distribution?  Note that the 
initial mass ratio distribution for our binary population is
flat\footnote{The cumulative initial mass ratio
distributions in Figs.~\ref{qrat_fractal_cumulative}
and~\ref{qrat_Plummer_cumulative} are not straight lines despite being
drawn from a flat mass ratio distribution. This is due to limiting the
lower mass of companions to be $> 0.1$\,M$_\odot$ meaning that M-dwarfs
in particular do not fill the entire range of possible mass
ratios. See Section~2.}.

In Figs.~\ref{qrat_fractal_cumulative}
and~\ref{qrat_Plummer_cumulative} we show the cumulative distributions
of mass ratios in the intermediate 62 -- 620\,au range after 1~Myr
of the fractal clusters
(Fig.~\ref{qrat_fractal_cumulative}) and the Plummer spheres
(Fig.~\ref{qrat_Plummer_cumulative}).  In both Figs.\,\,panel (a) shows
all ten realisations, and panel (b) shows the two most extreme
distributions.  The initial mass ratio distributions are shown by the
thick dashed grey lines (for the same reasons as in the separation
disributions each Plummer sphere has a {\em slightly} different
initial distribution).

Interestingly in the fractal clusters, the spread in the mass ratio
distributions is rather low (Fig.~\ref{qrat_fractal_cumulative}), and
even the two extremes look very similar (panel (b)).  A KS test on
these distributions fails to distinguish them.

However, in the
Plummer spheres where processing has been much more extreme we see
that the difference between the two extremes is greater than in the fractal clusters, but still not 
enough to be significantly different in a KS test.

Finally, we note that both the fractals and Plummer spheres with the two 
extremes in the final mass ratio
distribution are {\em not} the same clusters with the extremes in the
separation distribution.

Whether an encounter is destructive depends on the
distance of that encounter and the binding energy of the binary: a
closer binary requires a closer encounter to destroy it.  Binding 
energy also depends on the mass ratio, but these results suggest that the
most important factor is the distance of the encounter (which depends
on density but is stochastic with regards to the mass ratio of the
system).  In the case of the Plummer sphere clusters
(Fig.~\ref{qrat_Plummer_cumulative}(b)) both clusters have {\em more}
$q<0.4$ systems after processing than before (inspection of
Fig.~\ref{qrat_Plummer_cumulative}(a) shows that this is not always
the case).  One extreme has stayed fairly close to the initial mass
ratio distribution (the upper extreme), whilst the lower extreme has
evolved to have far more very high mass ratio systems ($q>0.9$).

\section{Discussion}
\label{discuss}

From the results presented in Section~\ref{results} we can see that
stochasticity in the destruction of intermediate binaries can have a
number of important consequences.

As would be expected, intermediate binary destruction 
depends on density.  Dense environments are far more effective at
destroying intermediate binaries.  Indeed, the definition of what is a
hard, soft, or intermediate binary depends on density -- a hard binary
in a very low-density environment could be an intermediate binary in
another environment.

However, the range of (visual) binary separations in clusters 
normally probed by observations of 10s -- 100s~au covers the 
intermediate binary regime in the range of densities found those
clusters of $10 - 10^5$ stars pc$^3$ (see King et al. 2012).
Therefore, {\em we almost always observe intermediate binaries 
in star forming regions.}

As would be expected, we find that the number of binaries
processed depends on the density. But this processing is always 
stochastic with 18 -- 42 per cent of intermediate binaries processed in
the relatively low-density fractals, to 37 -- 71 per cent in the
higher-density Plummer spheres.  Therefore, {\em the number of
intermediate binaries that are destroyed depends on density, but
with a stochastic factor of around 2 in the number destroyed.}

A key result is illustrated in Figs.~\ref{reipurth_fractal_cumulative}
and~\ref{reipurth_Plummer_cumulative} that {\em whatever the density, 
the initial intermediate separation distribution can be 
significantly altered}.  Wider
intermediate binaries are generally destroyed in preference to closer
intermediates, but this is not a strong relationship. For an initially substructured cluster, differences in
processing can easily be extreme enough to show very strong
statistical differences with e.g. a KS test. We emphasize here that only two different pairs of simulations resulted in significantly different separation distributions. However, we have no way of telling 
whether the clusters we observe in reality are themselves stochastic outliers.

Interestingly, as shown in Figs.~\ref{qrat_fractal_cumulative}
and~\ref{qrat_Plummer_cumulative} the change in the initial mass ratio
distributions are not as strong as in the separation distributions.
It seems that {\em intermediate binary destruction does not care about
the system mass ratio}.  That intermediate binary destruction is
insensitive to mass ratio means that the processed mass ratio
distributions are statistically the same as the initial mass ratio
distribution.  Therefore, {\em in clusters with low levels of
intermediate binary destruction, the mass ratio distribution should
reflect the initial mass ratio distribution}.  However, 
one has to know that there has been little processing to use this
result. 

These results have interesting implications for the interpretation of both
observations and numerical simulations. 

Clearly, when we observe a cluster we are observing a single
realisation of its initial binary population and the processing of
that population.  When we compare two clusters we compare two
different realisations and as we have seen, any two realisations may be
very different, even if their initial conditions were the same.

Further, as we have seen the dynamical processing of intermediate
binaries does depend on density.  It is important to note that
processing occurs very quickly and so it is not the current density
that is important, but the maximum density reached at some point in
the past that determines how effective binary processing is 
\citep[see also][]{Goodwin10, Parker09a, Parker11c}.

The problem we face when attempting to compare two intermediate binary
populations and determine if they come from the same initial
populations is therefore twofold.  Firstly, we need to have
information about the past state of the cluster to determine what 
level of processing we might expect {\em on average}.  Secondly, we 
have to account for the stochasticity in intermediate binary
destruction.  

Let us take two ONC-like clusters as an example.  We started with 106
binaries in the observed 62--620~au range.  (Actually, randomly sampling from
the same underlying distribution would give a range of 90 -- 110
binaries initially in that range).  We then find that 37 -- 71 per cent
of these can be destroyed in a dense cluster.  Therefore in numbers of
binaries alone, statistically the same initial population could result
in between 39 and 65 binaries remaining in that range -- a difference
of over a factor of two from random chance alone.  And this is before
we consider the possibility that these two populations have also evolved to 
statistically different separation and mass ratio distributions.

If we were to observe two clusters with similar masses and densities
and find that they had statistically very siginificant differences in
the numbers of binaries, and the separations and mass ratios of these
binaries, then we might reasonably conclude that we were looking at two
different initial populations and therefore a difference in how the
stars were formed.  However, as we have seen, that is sadly not the
case.  

Great care must also be taken when comparing simulations with 
observations. It is standard
procedure to average together the outcome of, say, ten simulations and
compare  the separation distribution (with standard deviation) to the
observed one
\citep[e.g.][]{Kroupa99,Parker09a,Parker11c,King12a}. However, the
observed distribution may in fact be an outlier and a failure to fit
the observed distribution might not mean that the model is `wrong',
conversely, a good fit to the observed distribution does not mean
the model is `right' (correctly fitting an outlier would be worrying
unless one's simulation was also an outlier in the same way).  
Ensembles of simulations are crucial to at
least determine a reasonable tolerance for the model, however nothing
will ever be able to determine if the observed realisation is an
outlier or not.

In future papers we will examine the observed binary separation
distributions in clusters from \citet{King12a} in light of these
results, we will also examine what separation ranges are of use in
distinguishing differences or otherwise in the star formation in
different regions.

\section{Conclusions}
\label{conclude} 

We have conducted $N$-body simulations of ONC-like clusters containing 1500
stars (750 primordial binary systems) in which we have kept the
initial binary
population constant, but varied the positions and velocities of the
systems  within ten realisations of the same cluster. We
have studied two different cluster morpholgies; a fractal cluster
undergoing cool collapse \citep[e.g.][]{Allison10} and a Plummer
sphere in virial equilibrium  \citep[e.g.][]{Parker09a}. We have
compared the intermediate separation distribution (62 -- 620\,au) in
these clusters to examine the importance of stochasticity in
intermediate binary destruction.

We conclude the following:

\noindent (i) The numbers of intermediate systems destroyed in
clusters can vary by a factor of two.

\noindent (ii) The separation distributions of intermediate systems 
in substuctured clusters can be altered such that they are statistically significantly  
different after just 1~Myr.

\noindent (iii) The mass ratio distributions change less than the
separation distributions, especially in low-density environments.

The results imply that the intermediate binary separation
distribution, which is the range most often observed in young
clusters, should be treated with caution when used to interpret the
dynamical history of a star cluster.  Even with a knowledge of
the initial conditions and probable dynamical history of a cluster,
stochasticity in intermediate binary destruction can very significantly
alter the initial population. Whilst most clusters evolve in a `typical' way, statistically significant outliers are not uncommon and we have no way of knowing if a single observed cluster is unusual because of differences in the initial conditions, or through a slightly unusual dynamical evolution.

\section*{Acknowledgements}

We thank the anonymous referee for a prompt and helpful review. The simulations in this work were performed on the \texttt{BRUTUS}
computing cluster at ETH Z{\"u}rich.

\bibliographystyle{mn2e}
\bibliography{general_ref}

\label{lastpage}

\end{document}